\documentclass{article} 

\usepackage{jheppub} 

\usepackage{bbm} %blackboard font for numbers
\usepackage{bm} %boldface
\usepackage{hyperref}
\usepackage{amsmath,amssymb,amsthm} %operations, symbols, theorem-proof style text
\usepackage{tensor} %tensor indices
\usepackage{braket} %dirac braket
\usepackage[utf8]{inputenc} %languages
\usepackage{parskip}
\usepackage{mathrsfs}
\usepackage{slashed}

\allowdisplaybreaks % pagebreak through equations
\hypersetup{
	colorlinks=true,
	allcolors=[rgb]{0.28, 0.24, 0.75}} %blue links and citations
\newcommand{\dd}{\text{d}}
\newcommand{\DD}{\text{D}}
\newcommand{\pp}{\partial}
\newcommand{\mpl}{M_\text{Pl}}
\newcommand{\meff}{m_\text{eff}}

\renewcommand{\[}{\left[}
\renewcommand{\]}{\right]}
\renewcommand{\(}{\left(}
\renewcommand{\)}{\right)}

\renewcommand{\epsilon}{\varepsilon}

\title{On the Aretakis Instability of Extremal Black Branes}

\author[a]{Calvin Y.-R. Chen}
\author[b]{\& \'Aron D. Kov\'acs}

\affiliation[a]{Abdus Salam Centre for Theoretical Physics,
Imperial College London, SW7 2AZ, United Kingdom; \\ 
Leung Center for Cosmology and Particle Astrophysics, Taipei 10617, Taiwan; \\
Center for Theoretical Physics, National Taiwan University, Taipei 10617, Taiwan}
\affiliation[b]{School of Mathematical Sciences, Queen Mary University of London, Mile End Road, E1 4NS London, UK}
\date{\today}

\emailAdd{cyrchen@ntu.edu.tw}
\emailAdd{a.kovacs@qmul.ac.uk}

\abstract{
We investigate how the Aretakis instability affects non-dilatonic extremal black $p$-branes by focusing on their near-horizon geometry.
Crucially, the strength of the instability, \textit{i.e.} the number of transverse derivatives needed to see non-decay/blow-up of fields on the horizon at late null time, is given by the scaling dimensions with respect to the near-horizon $\mathrm{AdS}_{p+2}$-factor.
This renders the problem of determining the severity of the Aretakis instability equivalent to computing the Kaluza--Klein spectrum of fields on Freund--Rubin spaces.
We use this to argue that non-dilatonic extremal black branes suffer from the Aretakis instability even in the absence of additional fields --- we find that this is weaker than for extremal black holes.
We also argue that the scaling dimensions determine the smoothness of stationary deformations to the original black brane background --- here, our findings indicate that generically more modes can lead to worse curvature singularities compared to extremal black holes.
}

\begin{document}

\maketitle

\section{Introduction}

Horizons of black holes are known to possess various intriguing features.
For example, in $D=4$ spacetime dimensions, uniqueness theorems dictate that all asymptotically flat black holes in the electro-vacuum fall within the Kerr--Newman family of solutions, which generically possess two horizons \cite{Heusler:1996jaf, Chrusciel:2012jk}.
In higher-dimensional spacetimes, uniqueness theorems are less restrictive \cite{Hollands:2012xy} --- infamously, in $D=5$ spacetime dimensions one can have black holes and black rings (with different horizon topology) with degenerate charge, mass, and angular momenta \cite{Emparan:2001wn}. 
Moreover, one can also consider objects with non-compact horizons and charged under higher-form fields, which universally arise within the bosonic sector of the low-energy effective theory of string theory or supergravity. 
Such extended objects with $p+1$-dimensional worldvolumes are known as black $p$-branes --- these play a crucial role as non-perturbative/solitonic objects in supergravity and string theory.
In all of the cases mentioned above, there exist sub-classes of solutions for which the global {asymptotic} charges parameterising the families saturate certain inequalities known as Bogomol'nyi–-Prasad-–Sommerfield (BPS) bounds, for which the horizons degenerate (see \textit{e.g.} \cite{Kunduri:2013gce} and references therein).
These are said to be extremal.

Since extremal black hole/brane solutions comprise a codimension-$1$ subset of the space of stationary solutions, one could expect that generic perturbations of extremal black holes/branes will eventually settle down to a non-extremal solution and thus the phenomenological significance of extremal solutions is unclear. 
Moreover, for astrophysical black holes, it was argued that there is an upper limit (known as the Thorne limit) on the angular momentum per unit mass that can be acquired by accretion \cite{Thorne:1974}. 
The final nail in the coffin is the third law of black hole mechanics, which forbids the formation of extremal black holes in finite time. 
It has, however, recently been shown that the third law of black hole mechanics does not hold in Einstein-Maxwell-charged scalar field theory: a Schwarzschild black hole can be ``charged up" with infalling charged matter and can evolve to an extremal Reissner-Nordst\"om (RN) black hole in finite null time \cite{Kehle:2022uvc}. (Nevertheless, the third law {\it does} hold for supersymmetric black holes \cite{Reall:2024njy,McSharry:2025iuz}.)
In fact, many observed compact objects believed to be black holes are close to extremality --- see \textit{e.g.} \cite{McClintock:2006xd, Gou:2011nq, Risaliti:2013cga, Miller:2013rca}, or \cite{Bambi:2019xzp} for a review. 

All these arguments aside, extremal objects are also useful to study from a theoretical perspective, as they are analytically more tractable.
For instance, a widely celebrated result in string theory is the microstate counting of $D=5$ RN black holes via the D1-D5 system \cite{Strominger:1996sh}.
A particularly important property of extremal black holes and branes for this work is the rigidity of their near-horizon geometry \cite{Kunduri:2013gce}.
For instance, the near-horizon geometry of an extremal black hole in $D$ dimensions is described by the Robinson-Bertotti solution $\mathrm{AdS}_{2} \times S^{D-2}$ \cite{Robinson:1959ev, Bertotti:1959pf} --- these are special cases of Freund-Rubin (FR) spaces $\mathrm{AdS}_{p+2} \times S^{D-p-2}$ \cite{Freund:1980xh}, which arise as the near-horizon geometry of certain extremal black branes. 
Another curious feature of extremal black holes, which we will be particularly interested in, is that they suffer from a classical linear instability known as the \textbf{Aretakis instability} \cite{Aretakis:2011ha, Aretakis:2012ei, Lucietti:2012sf, Lucietti:2012xr,Angelopoulos:2018yvt,Hadar:2017ven}\footnote{This may seem surprising. 
It is often said that supersymmetry implies dynamical stability --- the fact that this is too na\"ive was emphasised in \cite{Lucietti:2012xr}.
As they point out, saturation of a BPS bound is neither a necessary nor sufficient condition for dynamical stability. 
The positive-mass theorem does not imply full non-linear stability of Minkowski space \cite{Christodoulou:1993uv} and empty Anti-de Sitter (AdS) space is unstable against arbitrarily small perturbations despite being a supersymmetric solution in theories of supergravity \cite{Bizon:2011gg}.}.
This instability can be demonstrated (for massless fields in four spacetime dimensions) by constructing quantities for linear perturbations that are conserved along the null generators of the horizon of an extremal black hole. These conservation laws can then be used to show that sufficiently many transverse derivatives of the perturbations will grow polynomially in null time.

It is interesting to ask whether extremal black branes also suffer from an Aretakis-type instability. In the seminal works \cite{Gregory:1993vy,Gregory:1994bj}, it was shown that {\it sub-extremal} black branes exhibit a classical instability known as the Gregory-Laflamme (GL) instability. 
This is triggered by sufficiently long wavelength $s$-wave perturbations which, at the non-linear level, leads to the formation of a self-similar sequence of black holes connected by black string segments \cite{Lehner:2010pn,Lehner:2011wc,Figueras:2022zkg}. 
However, \cite{Gregory:1994tw} argued that the same mechanism does {\it not} lead to an instability for extremal black branes. 
Nevertheless, since the Aretakis instability is not related to the worldvolume dynamics (unlike the GL instability), one might still expect extremal black branes to suffer from a similar instability. In fact, this expectation was confirmed in \cite{Cvetic:2020axz} for linear scalar waves on black branes with $D-p=4$.

In this work, we will use symmetry arguments motivated by \cite{Gralla:2018xzo,Chen:2017ofv,Chen:2018jed} to derive the late-time behaviour of derivatives of fields on the horizon of generic non-dilatonic extremal black branes.
This allows us to clarify the origin of the Aretakis instability and generalise the discussion in \cite{Cvetic:2020axz} to generic $(D,p)$.
Further, we will use this to demonstrate the Aretakis instability for non-dilatonic extremal black branes without introducing additional fields.

In particular, we will argue that for a field $\phi$ with scaling dimension $\Delta$ on $\mathrm{AdS}_{p+2}$ within the near-horizon geometry of the extremal black $p$-brane, its derivatives along the near-horizon coordinate $\rho$ evaluated on the horizon will scale as 
\begin{equation}
    \pp_{\rho}^{n} \phi \big|_{\rho = 0} \sim v^{n-\Delta}
    \label{eq: aretakis scaling}
\end{equation}
at {large null time $v$}.
This shows that at least  $\lceil \Delta \rceil$ or $\lfloor \Delta \rfloor + 1$ transverse derivatives are necessary to see non-decay or blow-up respectively in null time\footnote{Note that these differ only when $\Delta$ is exactly integer.}.
In the case of $p=0$, this scaling for generic non-integer $\Delta$ was confirmed numerically in \cite{Lucietti:2012xr}.
The dynamical fields in the black branes and their corresponding near-horizon geometries, \textit{i.e.} the metric and form field, are naturally subject to perturbations.
We will show that these perturbations themselves trigger the Aretakis instability for non-dilatonic extremal black branes, even in the absence of additional fields.
This is determined by the Kaluza--Klein (KK) spectrum of these perturbations compactified over the $S^{D-p-2}$ in the near-horizon geometry.
For magnetically charged extremal black branes, this spectrum was previously computed in \cite{DeWolfe:2001nz, Bousso:2002fi, Kinoshita:2009hh, Brown:2013mwa, Hinterbichler:2013kwa}.
We extend this computation to electrically charged backgrounds for completeness, by using the Scalar-Vector-Tensor (SVT) decomposition from \cite{Kodama:2000fa, Kodama:2003jz, Kodama:2003kk} to find the decoupled wave equations governing a set of gauge-invariant master variables.
In both cases, masses of all perturbations lie above the Breitenlohner-Freedman (BF) bound, with gravitational vector and scalar modes sometimes saturating this.  
These correspond to the perturbations with smallest possible scaling dimensions $\Delta_{\text{min}} = (p+1)/2$.
From \eqref{eq: aretakis scaling}, we conclude that more non-compact horizon dimensions are associated to a weaker Aretakis instability --- more transverse derivatives are needed to see non-decay/blow-up of fields on the horizon at late times. 

The scaling dimensions of the perturbations also determine the behaviour of static \textbf{deformations to the near-horizon geometry} of extremal black branes.
For extremal black holes, the perspective taken in \cite{Horowitz:2022mly} and subsequent works \cite{Horowitz:2022leb, Horowitz:2024kcx} is that these describe the near-horizon geometries of deformations from the original extremal black holes.
The perturbations themselves (and hence the curvature associated to the deformed geometry) exhibit power-law scaling 
\begin{equation}
    \phi \sim \rho^{\Delta - (p+1)}
\end{equation}
near the horizon, and imply non-smoothness of the horizon.
When $\Delta < p+1$ this can lead to a blow-up of curvature invariants on the horizon.
This is the case for a multiple mixed gravitational/form field perturbations, although is most severe for the modes saturating the BF bound mentioned above.
These are understood to be physical, as they arise as the extremal limit of regular sub-extremal deformations.
As identified in \cite{Horowitz:2023xyl} for the extremal black holes, the extremal black branes also have deformations with $\Delta = p+1$ which exhibit marginal near-horizon scaling: Curvature invariants appear to be constant as the horizon is approached, but small deviations from $\Delta = p+1$ can cause curvature singularities.  
These are of particular interest as they appear to be UV sensitive.

\subsection*{Organisation}

The rest of this manuscript is organised as follows.
In section \ref{sec: extremal black branes}, we review non-dilatonic, electrically and magnetically charged, extremal black $p$-branes in $D$ dimensions and their near-horizon geometries in various coordinates.
Then, in section \ref{sec: aretakis instability}, we derive the scaling \eqref{eq: aretakis scaling} for general $(D,p)$, relying on the symmetries of the near-horizon geometry.
An equivalent derivation of this in alternative coordinates is given in appendix \ref{app: alternative deriv}.
To demonstrate that this is an instability of the background geometry even in the absence of additional fields, we compute the Kaluza--Klein spectrum for perturbations in section \ref{sec: perturbing freund-rubin}.
Finally, we relate these computations to the near-horizon scaling of deformations to undeformed extremal black branes in section \ref{sec: deformations of branes}.
Some of the subtleties related to this discussion are postponed to appendices \ref{app: multi-brane} and \ref{app: extremal limit}.

\subsection*{Conventions}

We will use units in which $\hbar= c = 1$ and metrics with mostly-plus signature $(-,+,\dots +)$.
Furthermore, we will use capital letters from the beginning of the Roman alphabet $A,B,\dots$ to denote indices on the full spacetime.
We will be interested in $p$-branes, which have $d=p+1$-dimensional worldvolumes, and we define $\tilde{d} = D-d-2$.
We will use letters from the beginning of the Greek alphabet $\alpha,\beta,\dots$ and capital letters from the middle of the Roman alphabet $I,J,\dots$ to denote indices of tensors on or orthogonal to the worldvolume respectively.
It is sometimes useful to single out a timelike coordinate on the worldvolume --- in that case, we will use lowercase letters from the beginning of the Roman alphabet $a, b, \dots$ to denote indices in the remaining spatial directions.
For global coordinates on the worldvolume, it will also be useful to introduce an additional component subject to a constraint --- we will denote indices in these directions using hatted lowercase letters from the beginning of the Roman alphabet, \textit{i.e.} $\hat{a}, \hat{b},\dots$. 
Euclidean vectors transverse to the worldvolume of the branes are denoted by boldface font. 
We will also use spherical coordinates in the directions transverse to the worldvolume, which we parameterise by $(\hat{\rho},x^{i})$, with indices in the spherical directions denoted by lowercase letters from the middle of the Roman alphabet $i,j,\dots $.
At this point, we warn the reader that various other radial variables $\rho$, $r$, $\lambda$, $\hat{r}$, and $\rho_{*}$ are used and defined throughout this work.
In the near-horizon geometry, $x^{\alpha}$ and $\rho$ will combine to coordinates on $\mathrm{AdS}$ --- we will use letters from the middle of the Greek alphabet $\mu,\nu,\dots$ to denote indices in those directions.
Covariant derivatives on the full spacetime, and separately on just AdS and the unit sphere will be denoted by $\nabla$, and $\DD$ and $\hat{\DD}$ respectively.
Similarly, the Laplace-Beltrami operator on the full spacetime or on just the AdS$_{n}$ factor, and the Laplacian on $S^{m}$ are denoted with $\Box$, $\Box_{\mathrm{AdS}_{n}}$, and $\hat{\Delta}_{m}$ respectively.
Finally, we will denote the Lie derivative with respect to a vector $K$ by $\pounds_{K}$.

\section{Extremal Black Branes and Freund--Rubin Compactifications \label{sec: extremal black branes}}

In this section, we will discuss the background solutions we are interested in.
We start by briefly reviewing black branes as solutions to Einstein gravity coupled to a form field in section \ref{subsec: black branes}. 
In section \ref{subsec: freund-rubin} we then discuss FR geometries --- the direct of product of AdS and a sphere --- which are the near-horizon geometry of extremal non-dilatonic charged black branes. 

\subsection{Black Branes \label{subsec: black branes}}

Let us start by reviewing our set-up for extremal black branes.
This is mostly based on \cite{Stelle:1996tz}, but see also \cite{Gibbons:1987ps, Stelle:1998xg, Horowitz:1991cd, Gibbons:1994vm, Lu:1995cs, Tseytlin:1996zb, vanderSchaar:1999tx} and references therein.

Consider the universal, non-dilatonic bosonic part of the effective action arising from string/supergravity theories
\begin{equation}
	S = \frac{\mpl^{D-2}}{2} \int \dd^{D}x \sqrt{-g} \[R -  \frac{1}{2n!} F_{(n)}^{2}\], 
	\label{eq: action} 
\end{equation}
where $\mpl$ is the Planck mass (related to the Gravitational constant), $F_{(n)} = \dd A_{(n-1)}$ is the field strength of an $n-1$-form gauge field, and $F_{(n)}^{2}  = F_{A_{1} \dots A_{n}}F^{A_{1} \dots A_{n}}$\footnote{{Note that, for convenience, we have taken $F$ not to be canonically normalised.}}.
The background equations of motion are
\begin{subequations}
	\label{eq: background eqs}
	\begin{align}
		&\nabla_{A_{1}}F^{A_{1}\dots A_{n}} = 0, \\
		&G_{AB} - \frac{1}{2(n-1)!}\[F\indices{_{A A_{2}\dots A_{n}}}F\indices{_{B}^{A_{2}\dots A_{n}}} - \frac{1}{2n}g_{AB}F^{2}\] =0.
	\end{align}
\end{subequations}
We are interested in $p$-branes, which have $d=p+1$-dimensional worldvolumes with isometry group
\begin{equation}
	\mathrm{ISO}(1,d-1) \times \mathrm{SO}(D-d) \subset \mathrm{ISO}(1,D-1).
\end{equation}
They are described by the metric
\begin{subequations}
    \label{eq: extremal black brane}
    \begin{align}
        &\dd s^{2} = H^{-2/d} \eta_{\alpha\beta}\dd x^{\alpha} \dd x^{\beta} + H^{2/\tilde{d}} \delta_{IJ}\dd y^{I} \dd y^{J}, \\
        &H(\hat{\rho}) = 1 + \(\frac{r_{0}}{\hat{\rho}}\)^{\tilde{d}},\quad  \hat{\rho} = |\mathbf{y}|,
    \end{align}
\end{subequations}
where $\alpha,\beta,\dots \in \{0,1,\dots,d-1\}$ and $I,J,\dots \in \{d,\dots,D-1\}$ denote worldvolume and transverse indices respectively, and we have defined $\tilde{d} = D-d-2$.
We can distinguish between elementary and solitonic black branes, which are charged electrically or magnetically under the form field respectively.
The non-zero components of the corresponding field strengths are 
\begin{subequations}
    \begin{alignat}{3}
		& \text{electric: } & \quad & F_{I \alpha_{1}\dots \alpha_{n-1}} = \sqrt{\frac{2(D-2)}{d\tilde{d}}} \epsilon_{\alpha_{1}\dots \alpha_{n-1}} \pp_{I}H^{-1}, \quad & & n=d+1, \\
		& \text{magnetic: } & \quad & F_{I_{1} \dots I_{n}} = \sqrt{\frac{2(D-2)\tilde{d}}{d}}r_{0}^{\tilde{d}} \epsilon_{I_{1} \dots I_{n} J} \frac{y^{J}}{\hat{\rho}^{n+1}}, \quad & & n=D-d-1.
    \end{alignat}
    \label{eq: black brane field strength}
\end{subequations}
We will henceforth concentrate on the electrically charged case and take $n=d+1$ --- this comes without loss of generality, as they are related via electromagnetic duality (at least at the level of the background solution).
To switch to Schwarzschild-like coordinates, we take $r^{\tilde{d}}= \hat{\rho}^{\tilde{d}} + r_{0}^{\tilde{d}}$.
Then, the metric \eqref{eq: extremal black brane} becomes
\begin{equation}
    \dd s^{2} = f^{2/d} \eta_{\alpha\beta} \dd x^{\alpha} \dd x^{\beta} + f^{-2} \dd r^{2} + r^{2} \dd \Omega^{2}_{\tilde{d}+1},\quad f(r) = 1 - \(\frac{r_{0}}{r}\)^{\tilde{d}}.
    \label{eq: extremal black brane schwarzschild}
\end{equation}
This makes it manifest that the degenerate horizon is located at the codimension-1 surfaces defined by $r=r_{0}$.

\subsection{Freund--Rubin Compactifications \label{subsec: freund-rubin}}

The near-horizon limit of \eqref{eq: extremal black brane} is reached by keeping the leading-order terms in expansion around $\hat{\rho} = 0$. 
Taking $\hat{\rho} = r_{0} (\rho/L)^{d/\tilde{d}}$ with $L=\frac{d}{\tilde{d}}r_{0}$, we find that 
\begin{equation}
	\dd s_{\text{NH}}^{2} = \frac{\rho^{2}}{L^{2}} \eta_{\alpha\beta}\dd x^{\alpha} \dd x^{\beta} + \frac{L^{2}}{\rho^{2}} \dd \rho^{2}  + \dd \Omega^{2}_{D-p-2},
	\label{eq: ads x sph}
\end{equation}
which describes $\mathrm{AdS}_{p+2} \times S^{D-p-2}$ with $\mathrm{AdS}_{p+2}$ length $L$.
Naturally, this is also a background solution to \eqref{eq: background eqs} in its own right.

Let us review how this works precisely.
The Ansatz for FR compactifications introduced in \cite{Freund:1980xh} is to take the total $D$-dimensional spacetime to be a direct product $M = M_{d+1} \times M_{\tilde{d}+1}$ of Einstein manifolds and the field strength to be proportional to the volume form on $M_{d+1}$\footnote{For completeness, let us note that for the near-horizon geometry of magnetically charged/solitonic black branes, we would take the field strength to be proportional to the volume form on $M_{\tilde{d}+1}$.}, \textit{i.e.}
\begin{subequations}
	\label{eq: freund-rubin ansatz}
	\begin{align}
		& F_{\mu_{1} \dots \mu_{d+1}} = f \epsilon_{\mu_{1}\dots \mu_{d+1}}, \\
		& \dd s^{2} = g_{\mu\nu} \dd x^{\mu} \dd x^{\nu} + g_{ij} \dd x^{i} \dd x^{j},
	\end{align}
\end{subequations}
with 
\begin{equation}
	R_{\mu\nu} = \frac{1}{d+1}{}^{(d+1)}R\,g_{\mu\nu},\quad R_{ij} = \frac{1}{\tilde{d}+1}{}^{(\tilde{d}+1)}R\, g_{ij},
\end{equation}
and $\mu,\nu,\dots \in \{0,1,\dots , d\}$.
This is solved by taking
\begin{equation}
	{}^{(d+1)}R = - \frac{(d+1)\tilde{d}}{2(D-2)}f^{2},\quad {}^{(\tilde{d}+1)}R = \frac{d(\tilde{d}+1)}{2(D-2)}f^{2},
\end{equation}
so, assuming $f \in \mathbb{R}$, we have $M_{d+1} = \mathrm{AdS}_{p+2}$ and $M_{\tilde{d}+1} = S^{D-p-2}$, as expected.
In Poincar\'e coordinates, this is precisely described by the metric \eqref{eq: ads x sph}, and we may identify
\begin{equation}
	f^{2} = 2(D-2)\frac{d}{\tilde{d}} \frac{1}{L^{2}}.
\end{equation}
When we consider perturbations to these geometries in section \ref{sec: perturbing freund-rubin}, it will be convenient to denote covariant derivatives with respect to the AdS and the spheres separately by $\DD$ and $\hat{\DD}$ respectively. 

\subsubsection{Coordinate Charts \label{subsubsec: coord charts}}

There are several coordinate charts that will be useful for discussing the FR compactifications we are interested in. 

First, note that Poincar\'e coordinates $(x^{\alpha},\rho,\Omega)$ in \eqref{eq: ads x sph} are not regular near the Poincar\'e/black brane horizon $\rho=0$.
Fortunately, \textbf{Gaussian null coordinates} were constructed in \cite{Kunduri:2013gce, Cvetic:2020axz}.
Splitting $x^{\alpha} = (t,x^{a})$, we may define these coordinates $(v,\hat{x}^{a},\lambda,\Omega)$, with 
\begin{equation}
	\rho = \lambda \sqrt{1-\frac{k^{2}}{L^{2}}},\quad t = v + \frac{L^{2}/\lambda}{1- \frac{\hat{x}^{2}}{L^{2}}},\quad x^{a} = \frac{\hat{x}^{a}L/\lambda}{1- \frac{\hat{x}^{2}}{L^{2}}}.
\end{equation}
In this case, the metric \eqref{eq: ads x sph} becomes
\begin{equation}
	\dd s_{\text{NH}}^{2} = 2 \dd v\[\dd \lambda + \frac{\lambda}{L}h_{a} \dd \hat{x}^{a} - \frac{1}{2} \frac{\lambda^{2}}{L^{2}} F \dd v\] + K_{ab} \dd \hat{x}^{a} \dd \hat{x}^{b} + r_{0}^{2} \dd \Omega^{2}_{\tilde{d}+1},
\end{equation}
where
\begin{equation}
	h_{a} = -\frac{2}{1- \frac{\hat{x}^{2}}{L^{2}}} \frac{\hat{x}_{a}}{L},\quad F = 1- \frac{\hat{x}^{2}}{L^{2}},\quad K_{ab} = \frac{1}{1- \frac{\hat{x}^{2}}{L^{2}}}\(\delta_{ab} + \frac{1}{1- \frac{\hat{x}^{2}}{L^{2}}}\frac{\hat{x}_{a}\hat{x}_{b}}{L^{2}}\).
    \label{eq: gaussian null}
\end{equation}

\begin{figure}[t]
    \centering
    \includegraphics[width=0.5\linewidth]{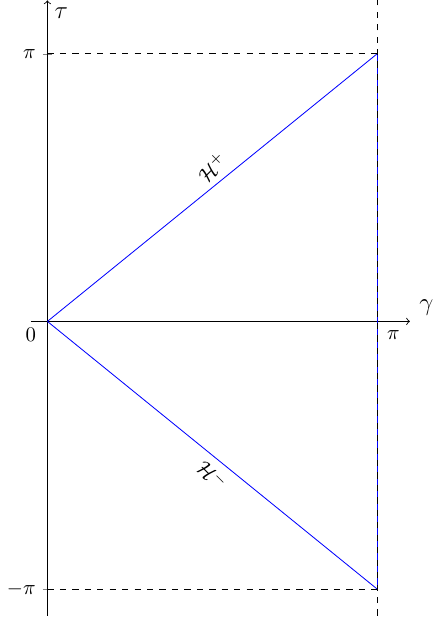}
    \caption{Global coordinates on AdS$_{p+2}$. 
    ${\cal H}^\pm$ denotes the future/past Poincar\'e horizons, $\gamma=\pi$ corresponds to right timelike infinity. 
    The region surrounded by the blue lines represents the exterior of (the near-horizon region of) the black brane, where we solve the wave equation.}
    \label{fig:ads2}
\end{figure}

It is often convenient to use a different affine parameter, $\hat{r} = \lambda\(1- \frac{\hat{x}^{2}}{L^{2}}\)$.
Additionally, since $\hat{x}^{2} <1$ we may also parameterise $\hat{x}^{a} = \tanh(\eta/L) \mu^{a}$ for $\mu_{a}\mu^{a} =1$.
Then, 
\begin{equation}
    \dd s^{2}_{\text{NH}} =  2 \dd v\cosh\(\frac{\eta}{L}\)^{2} \[\dd \hat{r}\,  - \frac{1}{2}\frac{\hat{r}^{2}}{L^{2}} \dd v\] + \dd \eta^{2} + L^{2}\sinh\(\frac{\eta}{L}\)^{2} \dd \Omega^{2}_{d-2} + r_{0}^{2} \dd \Omega_{\tilde{d}+1}^{2},
    \label{eq: warped product}
\end{equation}
with
\begin{equation}
	\rho = \hat{r} \cosh(\eta/L),\quad t = v + \frac{L^{2}}{\hat{r}},\quad x^{a} = \frac{L \tanh(\eta/L)}{\hat{r}} \mu^{a}.
\end{equation}
In this form, it is explicit that we may view $\mathrm{AdS}_{d+1}$ as a warped product $\mathrm{AdS}_{2} \times_{\eta} \mathbb{H}^{d-1}$.
This is crucial for the analysis in \cite{Cvetic:2020axz}.

For our symmetry argument for the Aretakis scaling in section \ref{sec: aretakis instability}, it will be useful to work in \textbf{global coordinates} $(\tau,\psi, y^{a},y^{p+1})$ on $\mathrm{AdS}_{p+2}$, where $y_{a} y^{a} + y_{p+1}y^{p+1}=1$ --- for this, it is convenient to bundle $y^{\hat{a}} = (y^{a},y^{p+1})$.
These coordinates are related to the Poincar\'e coordinates by
\begin{equation}
	t=L\,\frac{\sin\tau}{\cos\tau-y_{p+1} \cos\psi}, \quad \rho=L\,\frac{\cos\tau-y_{p+1} \cos\psi}{\sin\psi},\quad x^{a}=L\,\frac{y^{a}\cos\psi}{\cos\tau-y_{p+1} \cos\psi}.
\end{equation}
The relation to the Gaussian null coordinates may be expressed by introducing $\cos\gamma =y_{p+1}\cos\psi$:
\begin{equation}
	v=%L\,\frac{\sin\tau-\sin\gamma}{\cos\tau- \cos\gamma}=
	-\cot\frac{\tau+\gamma}{2}, \quad \hat{r}=L\,\frac{\cos\tau-\cos\gamma}{\sin\gamma},\quad \eta=L\cosh^{-1}\left(\frac{\sin\gamma}{\sin\psi}\right), \quad \hat x^{a}=\,\frac{y^{a}}{\sqrt{1-y_{p+1}^2}}.
\end{equation}
Note that the range of coordinates is $\tau\in [-\pi,\pi]$, $\gamma \in [0,\pi]$ such that $|\tau|<\gamma$ and correspondingly $\psi\in [0,\pi]$ (see also Fig. \ref{fig:ads2}). In particular, these conditions guarantee that $\rho>0$ and $\eta>0$.
In these coordinates, the near-horizon metric takes the form
\begin{equation}
    \dd s^2_{\rm NH}=\frac{L^2}{\sin^2\psi}\left(-\dd\tau^2+\dd\psi^2+\cos^2\psi\, \dd\Omega_p^2\right)+r_0^2 \dd\Omega_{D-p-2}^2.
	\label{eq: global coords}
\end{equation}

\section{The Aretakis Instability \label{sec: aretakis instability}}

We are now in a position to discuss the Aretakis instability.
As mentioned in the introduction, this is a classical linear instability which extremal black holes are known to suffer from.
This manifests itself in the polynomial growth of transverse derivatives of a massless field on the horizon of an extremal black hole in null time \cite{Aretakis:2011ha, Aretakis:2011hc, Aretakis:2012ei}.
The exponent characterising the growth can be shown to be related to the $\mathrm{AdS}_{2}$ scaling dimensions (which are given in terms of the effective masses) \cite{Lucietti:2012xr,Gralla:2018xzo}. In \cite{Cvetic:2020axz}, this was further generalised to a massive scalar field on a fixed non-dilatonic and electrically charged extremal black brane with $\tilde{d} = 2$. 

In this section, we will use symmetry arguments to derive the scaling in the Aretakis instability and generalise to generic massive fields with arbitrary $(D,p)$.
We will start by reviewing how this works for the extremal black holes (\textit{i.e.} when $p=0$) in section \ref{subsec: extremal bhs aretakis}, before generalising to extremal black branes in section \ref{subsec: extremal black branes aretakis}.

\subsection{Extremal Reissner-Nordstr\"om Black Holes \label{subsec: extremal bhs aretakis}}

We start by briefly reviewing the Aretakis instability for extremal RN black holes in $D=4$, following mostly \cite{Lucietti:2012xr, Hadar:2017ven}.

Consider the dynamics of a massless scalar field $\phi$ governed by the Klein-Gordon equation
\begin{equation}
    \Box_{g} \phi=0
\end{equation}
with the fixed background geometry \eqref{eq: extremal black brane schwarzschild} with $p=0$ as $g$. 
It is more convenient to work in ingoing Eddington-Finkelstein coordinates $(v,r,\Omega)$ with $\dd v = \dd t + \dd r / f^{2}$, in which the extremal RN metric \eqref{eq: extremal black brane schwarzschild} is given by
\begin{equation}
    \dd s^2=-f(r)^2 \dd v^2+2\dd v\, \dd r+r^2 \dd\Omega^2, \quad f(r)=\left(1-\frac{M}{r}\right),
    \label{eq: 4d rn ef}
\end{equation}
where $M = r_{0}$.
Expanding the scalar field in terms of spherical harmonics as
\begin{equation}
    \phi(v,r,\Omega)=\sum_{\ell,m} \phi_{\ell,m}(v,r)Y_{\ell,m}(\Omega),
\end{equation}
the wave equation reduces to
\begin{equation}
    2r\pp_{v}\pp_{r}\(r\phi_{\ell}\) +  \pp_{r}\[r^{2}f(r)^2 \pp_{r}\phi_{\ell}\] - \ell(\ell+1)\phi_{\ell} = 0,
    \label{eq: aretakis nh equation}
\end{equation}
where we were able to suppress the azimuthal label $m$ due to spherical symmetry.
Now we act with $\partial_r^n$ (where $n$ is a positive integer) on \eqref{eq: aretakis nh equation} and evaluate it on the horizon $r=M$. 
This gives
\begin{equation}
    2 \pp_{v} \pp_{r}^{n} \[r \pp_{r} \(r \phi_{\ell}\) \]\big|_{r=M} + \[n(n+1) - \ell(\ell+1)\]\pp_{r}^{n}\phi_{\ell}\big|_{r=M} =0.
    \label{eq: nh eq 4d rn}
\end{equation}
Taking $n=\ell$, we see that the quantity 
\begin{equation}
	H_{\ell} = M^{\ell-1} \pp_{r}^{\ell} \[r \pp_{r} \(r \phi_{\ell}\) \]\big|_{r=M}
\end{equation}
is conserved\footnote{{Note we have normalised $H_{\ell}$ to have the same mass dimension as $\phi_{\ell}$, in contrast to \cite{Lucietti:2012xr, Hadar:2017ven}.}}. 
A more detailed analysis \cite{Aretakis:2011ha,Aretakis:2011hc} shows that for $k\leq \ell$ the quantity $\partial^k\phi_{\ell}$ decays at late times ($v\to \infty$) on and off the horizon. 
It follows that 
\begin{equation}
	\lim_{v \rightarrow \infty} \pp_{r}^{\ell+1}\phi_{\ell} \big|_{r=M} = M^{-(\ell+1)} H_{\ell},
\end{equation}
\textit{i.e.} $\pp_{r}^{\ell+1}\phi_{\ell}$ does not decay at late times on the horizon. 
One can then show by induction that for $n \geq \ell+1$
\begin{equation}
	\pp_{r}^{n} \phi_{\ell}\big|_{r=M} \sim M^{-2n+\ell+1}H_{\ell}v^{n-\ell-1},\quad \text{as $v\rightarrow \infty$}
	\label{eq: 4d aretakis}
\end{equation}
\textit{i.e.} the $n$-th transverse derivative of $\phi_{\ell}$ grows polynomially in $v$ with a power $n-\ell-1$.

Note that conservation of $\pp_{\rho}^{\ell+1}$ follows from the near-horizon limit directly.
Taking the near-horizon limit of \eqref{eq: aretakis nh equation} directly gives the AdS wave equation in Poincar\'e coordinates
\begin{equation}
   2 \pp_{v}\pp_{\rho} \phi + \(\frac{\rho}{L}\)^{2}\pp_{\rho}^{2} \phi + \frac{2\rho}{L} \pp_{\rho}\phi - \meff^{2} \phi = 0,
\end{equation}
the $n$-th radial derivative of which, evaluated on the horizon $\rho=0$, is
\begin{equation}
    2 \pp_{v} \pp_{\rho}^{n+1} \phi + \(\frac{n(n+1)}{L^{2}}-\meff^{2}\) \phi = 0.
\end{equation}
Parameterising $\meff^{2} = \Delta^{(2)}(\Delta^{(2)}-1)$, this implies the conservation of $\pp_{\rho}^{\Delta^{(2)}}\phi$ when the AdS$_{2}$ scaling dimension $\Delta^{(2)}$ is integer. 
The blow-up of higher transverse derivatives was seen from the higher derivatives of the original wave equation. 
These are hidden in subleading terms in the near-horizon expansion, and are therefore missed by the near-horizon limit.

\subsubsection{Symmetry Argument}\label{sec:bh_aretakis_symmetry_arg}

Next, we present a new argument based on the symmetries of the near-horizon geometry (building on \textit{e.g.} \cite{Gralla:2018xzo,Chen:2017ofv,Chen:2018jed}) rather than conserved quantities\footnote{See also \textit{e.g.} \cite{Charalambous:2022rre,Charalambous:2024tdj,Charalambous:2025ekl} for related approaches in the context of computing Love numbers of various geometries.}. Our main motivation for doing this is that in a general extremal black brane geometry there appear to be no analogous conservation laws whereas the symmetry-based argument generalises to the case of black branes. Moreover, our argument captures the leading decay rates (rigorously established by \cite{Angelopoulos:2018uwb,Murata:2013daa}) of the scalar field just outside the horizon.

Since we are ultimately interested in the dynamics at late times near (and on) the horizon, it is useful to go into the near-horizon geometry $\mathrm{AdS}_2 \times S^{2}$, which has an enhanced isometry group $\mathrm{SL}(2,\mathbb{R})\times \mathrm{SO}(3)$, and work in global coordinates $\tau\in [-\pi,\pi]$ and $\psi\in(0,\pi]$ with $|\tau|<\psi$ introduced in section \ref{subsubsec: coord charts}.
For $p=0$, the metric reads as
\begin{equation}
    \dd s^2_{\text{NHRN}}=M^2\left[\csc^2\psi\, (-\dd \tau^2+\dd \psi^2)+ \dd\Omega^2\right].
\end{equation}
As before, we first expand the scalar field $\phi$ in terms of spherical harmonics, which gives the wave equation on $\mathrm{AdS}_2$ with unit radius 
\begin{equation}
   \biggl( -\partial_\tau^2+\partial_\psi^2-\frac{\ell(\ell+1)}{\sin^2\psi}\biggr)\phi_\ell=0. \label{eq:wave_AdS2_global}
\end{equation}
Note that from the spherical decomposition, $\ell$ is a non-negative integer. However, in the following argument, we will \textit{not} assume that $\ell$ is an integer --- it can be any non-negative real number. 
For reasons that will become obvious later, we will identify $\Delta^{(2)} = \ell+1$. 

To solve equation \eqref{eq:wave_AdS2_global}, it is useful to discuss the symmetries of the near-horizon geometry. The Killing vector fields in AdS$_2$ may be written as
\begin{subequations}
	\begin{align}
		&L_0=-i\,\partial_\tau, \nonumber \\
		&L_\pm=i \,e^{\pm i\tau}\,(-\cos\psi\, \partial_\tau\mp i\sin\psi\, \partial_\psi),
	\end{align}
	\label{eq: killing vectors global basis}
\end{subequations}
which obey the $\mathfrak{sl}(2,\mathbb{R})$ commutation relations
\begin{equation}
    [L_+,L_-]=-2L_0, \qquad [L_\pm,L_0]=\mp L_\pm.
\end{equation}
%
%We also note for later convenience that $\overline{L_\pm}=-L_{\mp}$ where the overbar denotes complex conjugation. 
Remarkably, the quadratic Casimir operator of $\mathfrak{sl}(2,\mathbb{R})$ turns out to be the wave operator (see \textit{e.g.} \cite{deWit:1999ui})
\begin{equation}
    {\cal C}\equiv \pounds_{L_0}\pounds_{L_0}-\frac12 \{\pounds_{L_-},\pounds_{L_+}\}=\Box_{\mathrm{AdS}_2},
\end{equation}
where $\pounds$ denote the Lie derivative. 
This observation allows us to construct basis functions that form a representation of the background geometry isometry group and expand solutions to the wave equation in this basis.

To see how this works, we first note that since ${\cal C}$ commutes with each generator of the isometry group, we may seek basis functions for solutions of \eqref{eq:wave_AdS2_global} as simultaneous eigenfunctions of ${\cal C}$ and one of the generators, \textit{e.g.} the dilatation generator $L_0$. 
That is, we will look for basis functions $\phi_{\ell,h}$ satisfying
\begin{subequations}
    \begin{align}
        & {\cal C}\phi_{\ell,h} =\ell (\ell+1) \phi_{\ell,h}, \label{eq: casimir_ev} \\
        & L_0\phi_{\ell,h} =h \phi_{\ell,h}. \label{eq: dilation_ev}
    \end{align}
\end{subequations}
The general solution to \eqref{eq: dilation_ev} is simply
\begin{equation}
    \phi_{\ell,h}=e^{ih\tau} F(\psi).
\end{equation}
Plugging this into \eqref{eq: casimir_ev} gives the ODE
\begin{equation}
    \frac{\dd^2 }{\dd\psi^2}F(\psi)+\left(h^2-\frac{\ell(\ell+1)}{\sin^2\psi}\right)F(\psi)=0. \label{eq:ads_2_psi}
\end{equation}
The general solution of this ODE may be written in terms of hypergeometric functions. 
Setting 
\begin{equation}
	F(\psi)=(\sin\psi)^{\ell+1}\,f\left(\sin^2\tfrac{\psi}{2}\right),
\end{equation}
the ODE \eqref{eq:ads_2_psi} reduces to the hypergeometric differential equation
\begin{equation}
    (1-z)\,z \, f''(z)+(c-(a+b+1)z)f'(z)-ab\,f(z)=0,\quad z = \sin^{2}\tfrac{\psi}{2}
\end{equation}
with
\begin{equation}
    a=\ell+1-h, \qquad b=\ell+1+h, \qquad c=\ell+\frac32\,.
\end{equation}
One solution is thus given by
\begin{equation}
    F(\psi)=\,(\sin\psi)^{l+1} ~{}_2 F_1\left(\ell+1-h,\, h+\ell+1,\,\tfrac32+\ell,~\sin^2\tfrac{\psi}{2}\right).
\end{equation}
This solution is regular at $\psi = 0$. 
The second independent solution turns out to be always divergent at $\psi = 0$ and we shall hence discard such solutions.

The differential equation \eqref{eq:ads_2_psi} is symmetric under $h \mapsto -h$, so we can choose $h \geq 0$ without loss of generality.
Regularity at $\psi = \pi$ now requires $h=\ell+1+n$ for $n\in\mathbb{N}_{\geq 0}$ and therefore we obtain a discrete set of regular solutions for the radial ODE
\begin{equation}
	\begin{aligned}
		F_n(\psi)&=\tilde c_{n,\ell}\,(\sin\psi)^{\ell+1} ~{}_2 F_1\left(-n,\, 2\Delta^{(2)}+n,\,\Delta^{(2)}+\tfrac12,~\sin^2\tfrac{\psi}{2}\right)\nonumber \\
		&=c_{n,\ell}\,(\sin\psi)^{\ell+1}\,P_n^{\left(\Delta^{(2)}-\tfrac12,\,\Delta^{(2)}-\tfrac12\right)}(\cos\psi),
	\end{aligned}
\end{equation}
where $P^{(\alpha,\beta)}_n$ denotes the Jacobi polynomials (see \textit{e.g.} \cite{Szegö:1959}) and $c_{n,\ell}$ are suitable normalisation constants chosen such that
\begin{equation}
    \int_{0}^{\pi}\dd\psi\, \left|F_n(\cos\psi)\right|^2=1.
\end{equation}
Note that the functions $F_{n}(\psi)$ form an orthonormal basis on the Hilbert space $L^2([0,\pi],\dd\psi)$ \cite{Szegö:1959}. 
Therefore, we obtained a set of basis functions
\begin{equation}
	\phi_{n,\ell,m}(\tau,\psi,\Omega)=c_{n,\ell}\,e^{i (\Delta^{(2)}+n)\tau}\,(\sin\psi)^{\Delta^{(2)}}\,\, P_n^{\left(\Delta^{(2)}-\tfrac12,\,\Delta^{(2)}-\tfrac12\right)}(\cos\psi) \,Y_{\ell,m}(\Omega)
\end{equation}
for the near-horizon wave equation \eqref{eq:wave_AdS2_global}. 
Introducing the standard Klein-Gordon inner product on a Cauchy slice $\Sigma$ with unit normal $n$
\begin{equation}
    (\phi_1,\phi_2)_\Sigma\equiv i\int\limits_{\Sigma} \star\, \iota_n(\overline{\phi_1} \dd\phi_2-\phi_2\dd \overline{\phi_1} ),
\end{equation}
one can easily check that on slices of constant $\tau$, the basis functions satisfy
\begin{equation}
    (\phi_{n,\ell,m},\phi_{n',\ell',m})_{\Sigma_\tau}=\delta_{nn'}\delta_{\ell\ell'}\delta_{mm'}.
\end{equation}

It is also interesting to note that (for fixed angular coordinates) these basis functions form the discrete series representation $\mathscr{D}^-_\ell$ of $SL(2,\mathbb{R})$. 
One can also generate these solutions by starting with the lowest weight element $n=0$ for each $\ell$: The function 
\begin{equation}
	\phi_{0,\ell}=c_{n,\ell}\,e^{-i (\ell+1)\tau}\,\sin^{\ell+1}\psi
\end{equation}
satisfies
\begin{equation}
	L_{-}\,\phi_{0,\ell}=0.
\end{equation}
The basis functions with $n>0$ can then be obtained by repeated actions of the raising operator $L_+$:
\begin{equation}
	\phi_{n,\ell}= (L_{+})^n\,\phi_{0,\ell}.
\end{equation}

To obtain the behaviour of the solutions on the future horizon (corresponding to $\rho=0$ and $\tau=\psi$), we note that
\begin{equation}
    \sin\psi=2\,(1+v^2)^{-1/2}\left[\rho^2+(2+\rho v)^2\right]^{-1/2}.
\end{equation}
From this it follows that on the horizon and for fixed $\rho >0$, we have
\begin{subequations}
	\begin{align}
		&|\phi_{n,\ell,m}|_{{\cal H}^+}(v)\sim v^{-\Delta^{(2)}}, && v\to\infty, \\
		&|\phi_{n,\ell}|_{\rho>0}(v)\sim v^{-2\Delta^{(2)}}, &&  v\to\infty,
	\end{align}
\end{subequations}
respectively, \textit{i.e.} the latter decay twice as fast as on the horizon. 
Since these decay rates hold for all the basis functions, they also hold for linear combinations of them, \textit{i.e.} for general solutions to the near-horizon wave equation that respect the boundary conditions imposed above. 
By taking radial derivatives of the basis functions on the horizon, we also recover
\begin{equation}
    |\partial_\rho^k\phi_{n,\ell,m}|_{{\cal H}^+}(v)\sim v^{k-\Delta^{(2)}}, \hspace{2cm} v\to\infty
\end{equation}
and hence, for $k>\Delta^{(2)}$, the quantities $|\partial_\rho^k\phi_{n,\ell,m}|_{{\cal H}^+}(v)$ will grow polynomially along the horizon.

%\begin{equation*}
 %   \phi_{n,\ell}|_{{\cal H}^+}(\tau)=e^{i (\Delta^{(2)}+n)\tau}\,(\cot\tau)^{-\Delta^{(2)}}\left(\cos\tau\right)^{\Delta^{(2)}+n}\,\, {}_2 F_1\left(\Delta^{(2)}+\frac12-\frac{n}{2},\Delta^{(2)}+\frac{n}{2},\Delta^{(2)}+\frac12,-\tan^2\psi\right)
%\end{equation*}
%At late times $\tau\to \pi$ the hypergeometric function approaches a constant and hence
%\begin{equation}
%    |\phi_{n,\ell}|_{{\cal H}^+}(\tau)\sim |\cot\tau|^{-\Delta^{(2)}}, \hspace{2cm} \tau\to\pi
%\end{equation}

%\begin{align}
 %   &\sin\psi=\frac{2}{\sqrt{(1+v^2)\left(\rho^2+(2+\rho v)^2\right)}}, \qquad \cos\psi=-\frac{\rho+v(2+\rho v)}{\sqrt{(1+v^2)\left(\rho^2+(2+\rho v)^2\right)}} \nonumber \\
  %   &\sin\tau=\frac{2(1+\rho v)}{\sqrt{(1+v^2)\left(\rho^2+(2+\rho v)^2\right)}}, \qquad \cos\tau=\frac{\rho-v(2+\rho v)}{\sqrt{(1+v^2)\left(\rho^2+(2+\rho v)^2\right)}} \nonumber
%\end{align}

\subsection{Extremal Black Branes \label{subsec: extremal black branes aretakis} }
%\label{sec:brane_aretakis}

We now consider the wave equation on an extremal black brane background. 
Since we are still primarily interested in the behaviour of the solution near the horizon at late times, it seems reasonable to work with the near-horizon geometry\footnote{This assumption seems reasonable as long as the solution to the wave equation decays on and outside the horizon in the full geometry. 
See also \cite{Gajic:2024uko} for a more rigorous work in this direction.}. 
Hence, we consider the AdS$_{d+1}$ wave equation\footnote{See also \textit{e.g.} \cite{Ishibashi:2003jd, Ishibashi:2004wx,Warnick:2012fi,Holzegel:2012wt,Holzegel:2013vwa,Holzegel:2015swa,Vasy2009TheWE} for rigorous studies on the wave equation in (asymptotically) AdS spacetimes.}, which in Poincar\'e coordinates takes the form
\begin{equation}
	\Box_{g}\phi = \Box_{\text{AdS}_{p+2}} \phi + \frac{1}{r_{0}^{2}}\hat{\Delta}_{D-p-2} \phi =0,\end{equation}
where 
\begin{equation}
	\Box_{\text{AdS}_{p+2}} \phi = \(\frac{L}{\rho}\)^{2} \pp_{\alpha}\pp^{\alpha} \phi + \frac{\rho^{2}}{L^{2}}\(\pp_{\rho}^{2}\phi + \frac{p+2}{\rho}\pp_{\rho}\phi \) 
	\label{eq: bg scalar eom} 
\end{equation}
and $\hat{\Delta}_{m}$ is the Laplacian on $S^{m}$.
For a field with mass $\meff$ propagating on a fixed $\mathrm{AdS}_{p+2}$ geometry, a natural quantity to define is the scaling dimension
\begin{equation}
	\Delta^{(p+2)}(\meff) = \frac{d}{2}\(1  + \sqrt{1 + \frac{4\meff^{2}}{d^{2}}}\).
	\label{eq: scaling dimension}
\end{equation}

To study the near horizon dynamics in black brane spacetimes, it is more suitable to consider global coordinates \eqref{eq: global coords}, summarised in section \ref{subsubsec: coord charts}.
In these coordinates, 
\begin{equation}
	\Box_{\text{AdS}_{p+2}}\phi = \frac{\sin^2\psi}{L^{2}}\left[-\partial_\tau^2\phi+\tan^p\psi\,\partial_\psi\left(\cot^p\psi\,\partial_\psi\phi\right)\right] + \frac{\tan^2\psi}{L^{2}} \hat{\Delta}_{p}\phi.
    \label{eq: global ads box}
\end{equation}
Once again, we may first decompose $\phi$ in terms of spherical harmonics on $S^{D-p-2}$:
\begin{equation}
    \phi=\sum_{\ell,\ell'} \phi_{\ell'}(\tau,\psi) Y_{\ell'}(\theta')
\end{equation}
where $\theta'$ are coordinates on $S^{D-p-2}$ and $\ell'$ is a multi-index for the spherical quantum numbers. In more detail, let $|\ell_1'|\leq \ell_2'\leq \ldots \ell_{D-p-2}'$. Then the spherical harmonics $Y_{\ell'_1\ldots \ell'_{D-p-2}}$ are eigenfunctions of the Laplacians
\begin{equation}
     \hat{\Delta}_{D-p-2} Y_{\ell'_1\ldots \ell'_{D-p-2}}=-\ell'_{D-p-2}(\ell'_{D-p-2}+D-p-3)Y_{\ell'_1\ldots \ell'_{D-p-2}}.
\end{equation}
With a slight abuse of notation, we will simply use the shorthand notation $\ell'\equiv \ell'_{D-p-2}$ from now on, unless otherwise stated.

We will now further decompose the scalar field using the symmetries of the near-horizon geometry. 
To this end, we now turn to a discussion of the Lie algebra of $\mathfrak{so}(p+1,2)$, the isometry group of AdS$_{p+2}$. 
In global coordinates, the Killing vector fields of AdS$_{p+2}$ can be written as
\begin{subequations}
	\begin{align}
		&L_0=-i\,\partial_\tau, \qquad \qquad L_{\hat{a}\hat{b}}=-i\,\left(y_{\hat a}{\partial_{\hat b}} -y_{\hat b}{\partial_{\hat a}} \right), \\
		&L_{\pm,\hat a}=i\, e^{\pm i\tau}\left[\left(\cos\psi\, \partial_\tau\pm i\sin\psi \,\partial_\psi\right)\, y_{\hat a}+\mp \,i \sec\psi\,\left(\delta_{\hat a}^{\hat b}-y_{\hat a} y^{\hat b}\right)\,\partial_{\hat b}\right],
	\end{align}
\end{subequations}
which satisfy the commutation relations
\begin{subequations}
	\begin{align}
		&[L_{\pm,\hat a},L_0]=\mp L_{\pm,\hat a}, \\
		&[L_{+,\hat a},L_{-,\hat b}]=-2(L_0 \eta_{\hat{a}\hat{b}}+i\,L_{\hat{a}\hat{b}}), \\
		&[L_{\hat{a}\hat{b}},L_{\pm,\hat c}]=i\,(\eta_{\hat{a}\hat{c}}L_{\pm,\hat b}-\eta_{\hat{b}\hat{c}}L_{\pm,\hat a}).
	\end{align}
\end{subequations}
Similarly to the case of AdS${}_2$, the quadratic Casimir operator of $\mathfrak{so}(p+1,2)$ can be written as
\begin{equation}
    {\cal C}_{2,{\rm AdS}_{p+2}}=\pounds_{L_0}\pounds_{L_0}-\{\pounds_{L_{-,\hat a}},\pounds_{L_{+,\hat a}}\}-\frac12 \pounds_{L_{\hat{a} \hat{b}}}\pounds_{L_{\hat{a}\hat{b}}}=\Box_{\text{AdS}_{p+2}}.
\end{equation}
Note that the generators $L_0$, $L_{\pm,\hat a}$ form an $\mathfrak{sl}(2,\mathbb{R})$ subalgebra. We may now look for basis functions that are simultaneous eigenfunctions of $L_0$, $L_{\hat{a}\hat{b}}L^{\hat{a}\hat{b}}\equiv\hat{\Delta}_{p}$ and ${\cal C}$:
\begin{subequations}
	\begin{align}
		&{\cal C}_{2,{\rm AdS}_{p+2}}\,\phi_{h,\ell,\ell'}=\ell'(\ell'+D-p-3) \frac{L^2}{r_0^{2}}\,\phi_{h,\ell,\ell'}, \\
		& \hat{\Delta}_{p}\,\phi_{h,\ell,\ell'}=-\ell(\ell+p-1) \,\phi_{h,\ell,\ell'}, \\
		& L_0\, \phi_{h,\ell,\ell'}=-h\, \phi_{h,\ell,\ell'}. 
	\end{align}
\end{subequations}
It follows that 
\begin{equation}
    \phi=\sum_{h,\ell,\ell'} e^{ih\tau}\chi_{h,\ell,\ell'}(\psi) Y_{\ell}(\theta)Y_{\ell'}(\theta'),
\end{equation}
where $\chi_{h, \ell,\ell'}(\psi)$ satisfies the ODE
\begin{equation}
	\tan^p\psi\, \frac{\dd}{\dd\psi}\left(\cot^p\psi\,\frac{\dd}{\dd\psi}\chi(\psi) \right)+\left(h^2-\frac{\mu_\ell^2}{\cos^2\psi}-\frac{\nu_{\ell'}^2}{\sin^2\psi} \right)\,\chi(\psi)=0,
    \label{eq: definition of nu}
\end{equation}
with $\mu_\ell\equiv\ell (\ell+p-1)$ and $\nu_{\ell'}^2\equiv \tfrac{L^2}{ r_0^{2}} \ell'(\ell'+D-p-3)$. 
This can be solved in terms of hypergeometric functions. 
One solution is given by
\begin{equation}
	\chi(\psi)=(\sin\psi)^{\Delta^{(p+2)}(\nu_{\ell})}(\cos\psi)^{\ell}\, {}_2 F_1\left[\frac{\Delta^{(p+2)}(\nu_{\ell})+\ell-h}{2},\frac{\Delta^{(p+2)}(\nu_{\ell})+\ell+h}{2},\ell+\frac{p+1}{2},\cos^2\psi\right].
	\label{eq:brane_chi_hyp}
\end{equation}
For $p>0$ and $\nu_{\ell'}^{2} \geq 0$, this solution is regular at the horizon $\psi = \pi/2$, while the second independent solution diverges at $\psi = \pi/2$\footnote{Note that $p=0$ is the case we discuss in section \ref{subsec: extremal bhs aretakis}, albeit using a different basis for the solutions.
In that case, the two independent solutions are
\begin{subequations}
    \begin{align}
        F_{1} &= \sin(\psi)^{\Delta}\,{}_{2}F_{1}\(\frac{\ell+1-h}{2},\frac{\ell+1+h}{2};\frac{1}{2};\cos(\psi)^{2}\) \\
        F_{2} &= \cos(\psi) \sin(\psi)^{\Delta}\,{}_{2}F_{1}\(\frac{\ell+2-h}{2},\frac{\ell+2+h}{2};\frac{3}{2};\cos(\psi)^{2}\),
    \end{align}
\end{subequations}
which are both regular at $\psi = \pi/2$.
Regularity at $\psi = 0$ and $\psi = \pi$ now imposes
\begin{equation}
    h = \ell + 1 +n
\end{equation}
with even $n \in 2\mathbb{N}_{\geq 0}$ for $F_{1}$ and odd $n \in 2\mathbb{N}_{\geq 0} + 1$ for $F_{2}$.
The second branch of solutions with odd $n$ becomes irregular at asymptotic infinity for $p >0$.}.

Imposing decay at infinity $\psi=0$ and $\psi = \pi$ requires $\frac12(\Delta^{(p+2)}(\nu_{\ell})+\ell-h)$ to be a non-positive integer, \textit{i.e.}
\begin{equation}
	h=\Delta^{(p+2)}(\nu_\ell)+\ell+2n, \qquad n=0,1,\ldots \, .
	\label{eq:brane_quant_cond}
\end{equation}
The resulting solutions 
\begin{equation}
    \chi_{n,\ell,\ell'}(\psi)=\tilde c_{n,\ell,\ell'} (\sin\psi)^{\Delta^{(p+2)}(\nu_{\ell})}(\cos\psi)^{\ell}\, {}_2 F_1\left[-n,\,\Delta^{(p+2)}(\nu_{\ell})+\ell+n,\,\ell+\frac{p+1}{2},\cos^2\psi\right]
\end{equation}
can now be expressed in terms of the Jacobi polynomials
\begin{equation}
     \chi_{n,\ell,\ell'}(\psi)= c_{n,\ell,\ell'} (\sin\psi)^{\Delta^{(p+2)}(\nu_{\ell})}(\cos\psi)^{\ell}\, P_n^{(\alpha_\ell,\beta_{\ell'})}(\cos 2\psi),
\end{equation}
with
\begin{equation*}
         \alpha_\ell\equiv \ell+\frac{p-1}{2}, \qquad  \beta_{\ell'}\equiv \Delta^{(p+2)}(\nu_{\ell'})-\frac{p+1}{2}.
\end{equation*}
It follows from the orthogonality properties of the Jacobi polynomials that the functions $\chi_{n,\ell,\ell'}$ form an orthonormal basis (with suitably chosen $c_{n,\ell,\ell'}$) on the Hilbert space $L^2([0,\pi],\dd\psi)$.

We have therefore constructed a set of basis functions which can now be labelled by $n$, $\ell$, and $\ell'$ and are given by
\begin{equation}
	\begin{aligned}
		\phi_{n,\ell,\ell'}=&c_{n,\ell,\ell'}\,\exp\left[i (\Delta^{(p+2)}(\nu_\ell)+\ell+2n)\right]\,(\sin\psi)^{\Delta^{(p+2)}(\nu_\ell)}(\cos\psi)^{\ell}\, \nonumber \\
    &\qquad \qquad\times P_n^{(\alpha_\ell,\beta_{\ell'})}(\cos 2\psi)\, Y_\ell(\theta)\,Y_{\ell'}(\theta').
	\end{aligned}
\end{equation}
These basis functions satisfy orthogonality conditions \textit{w.r.t.} the Klein-Gordon inner product
\begin{equation}
    \left(\phi_{\tilde n,\tilde \ell,\tilde \ell'},\phi_{n,\ell,\ell'}\right)_{\Sigma_\tau}=\delta_{\tilde n n}\delta_{\tilde\ell \ell}\delta_{\tilde \ell',\ell'},
\end{equation}
and form a discrete representation of the conformal group. 
The elements of this representation can be generated from the lowest-weight element, which is annihilated by all the lowering operators, \textit{i.e.} satisfies
\begin{equation}
    L_{-,\hat a}\,\phi_{0,0,\ell'}=0,
\end{equation}
and similarly for the lowering operators of the $\mathfrak{so}(p+1)$ when $p\geq 2$. 
The latter condition implies that $\phi_{0,0,\ell'}$ must be independent of the coordinates $y^{\hat a}$ on parameterising $S^{p}$. 
Then we have
\begin{equation}
	\left(\cos\psi\, \partial_\tau- i\sin\psi \,\partial_\psi\right) \phi_{0,0,\ell'}(\tau,\psi)=0
\end{equation}
which gives
\begin{equation}
    \phi_{0,0,\ell'}=e^{i h \tau}\,(\sin\psi)^h.
\end{equation}
Plugging this into the wave equation yields
\begin{equation}
	h=\Delta_\pm^{(p+2)}(\mu_\ell).
\end{equation}
Decay at infinity selects the positive exponent $h=\Delta_+^{(p+2)}(\mu_\ell)$. 
The basis functions with $\ell>0$ and $n>0$ can then be constructed by repeated actions of the raising operators. 
In particular, one may explicitly check that
\begin{equation}
    (L_{+,1}-i L_{+,2})^\ell \left(e^{i h\tau}\sin^h\psi\right)\propto\, e^{i(h+\ell)} (\sin\psi)^h(\cos\psi)^\ell (y_1-i\,y_2)^\ell,
\end{equation}
and hence
\begin{equation}
    (L_{+,1}-i L_{+,2})^\ell \phi_{0,0,\ell'}\propto \phi_{0,\ell,\ell'},
\end{equation}
with the SO$(p+1)$ labels explicitly given by $\ell_1=-\ell$, $\ell_2=\ldots=\ell_p=\ell$. 
For fixed $\ell$, the basis elements with various $\ell_i$ labels can then be generated by the repeated actions of the $\mathfrak{so}(p+1)$ raising operators. 
Finally, the label $n$ can be increased by the repeated action of
\begin{equation}
    (L_{+,\hat a}L_{+}^{\hat a})^n\phi_{0,\ell,\ell'}\propto \phi_{n,\ell,\ell'}.
\end{equation}

Now we shall turn to the discussion of the late-time dynamics on the horizon. Using the transformation rules, we can see that
\begin{equation}
    \sin\psi=2\,\left(\cosh \frac{\eta}{L}\right)^{-1}\left(1+\frac{v^2}{L^2}\right)^{-1/2}\left[\frac{\hat r^2}{L^2}+\left(2+\frac{\hat r v}{L^2}\right)^2\right]^{-1/2}
\end{equation}
Therefore, it follows that on the horizon the basis functions satisfy
\begin{subequations}
	\begin{align}	
		&|\phi_{n,\ell,\ell'}|_{{\cal H}^+}(v)\sim v^{-\Delta^{(p+2)}(\nu_{\ell'})}, && v\to\infty, \\
		&|\phi_{n,\ell,\ell'}|_{\rho>0}(v)\sim v^{-2\Delta^{(p+2)}(\nu_{\ell'})}, && v\to\infty
	\end{align}
\end{subequations}
on and off the horizon respectively. 
Therefore, we conclude that the late-time behaviour of the basis functions and thus general solutions is determined by the scaling exponent $\Delta^{(p+2)}(\nu_{\ell'})$. 
Taking derivatives of the basis functions \textit{w.r.t.} $\hat r$ gives
\begin{equation}
    |\partial_{\hat r}^k\phi_{n,\ell,\ell'}|_{{\cal H}^+}(v)\sim v^{k-\Delta^{(p+2)}(\nu_{\ell'})}, \hspace{2cm} v\to\infty,
\end{equation}
meaning that if $k>\Delta^{(p+2)}(\nu_{\ell'})$ then the $k^{\text{th}}$-order transverse derivatives of a generic solution will grow polynomially along the horizon.

\subsubsection{Aretakis Scaling}

Now, the AdS$_{p+2}$ wave equation for a field with generic mass $m$ takes the form
\begin{equation}
    \(\Box_{\mathrm{AdS}_{p+2}} - \meff^{2}\) \phi = 0,
    \label{eq: generic ads wave eq}
\end{equation}
which, by comparison with \eqref{eq: global ads box} and \eqref{eq: definition of nu} is equivalent to a shift in $\nu_{\ell'}^{2} \mapsto \nu_{\ell'}^{2} + L^{2}m^{2} \geq L^{2}\meff^{2}$.
The mode which saturates this inequality ($\ell'=0$) will be dominant for the Aretakis instability.
We can therefore conclude that, for any field satisfying an $\mathrm{AdS}_{p+2}$ wave equation of the form \eqref{eq: generic ads wave eq}, the $\mathrm{AdS}_{p+2}$ scaling dimension \eqref{eq: scaling dimension} is the quantity which determines the strength of the Aretakis instability.
This is captured by \eqref{eq: aretakis scaling}.
We will see that this is precisely the case for perturbations of the background itself in the next section.

\section{Perturbing Freund--Rubin \label{sec: perturbing freund-rubin}}

We would now like to establish whether intrinsic perturbations of non-dilatonic extremal black branes trigger the Aretakis instability, \textit{i.e.} whether such geometries suffer from the Aretakis instability even in the absence of additional fields.

We have argued that the strength of the Aretakis instability, \textit{i.e.} the number of transverse derivatives needed to see non-decay/blow-up in null time on the horizon, is determined by the KK spectrum of modes compactified on the sphere within the near-horizon geometry of the (non-dilatonic) extremal black branes.
On magnetically charged (or solitonic) branes, this was already determined in a series of works \cite{DeWolfe:2001nz, Bousso:2002fi, Kinoshita:2009hh, Brown:2013mwa, Hinterbichler:2013kwa}.
While the theory itself enjoys electromagnetic duality, which relates the electrically charged extremal black brane backgrounds to ones with magnetic charge, picking a background itself spontaneously breaks this symmetry --- this means that perturbations on an electrically charged extremal black brane will not be identical to those on the magnetically charged background. 
For completeness, we will therefore compute the KK spectrum of perturbations on electrically charged extremal black branes using the methods from \cite{Kodama:2000fa, Kodama:2003jz, Kodama:2003kk} in section \ref{subsec: perturbation equations}. The summary of the results and the discussion of their implications on the Aretakis instability can be found in section \ref{subsec: aretakis from kk}.

\subsection{Perturbation Equations \label{subsec: perturbation equations}}

We take the FR geometry \eqref{eq: freund-rubin ansatz} (\textit{i.e.} the near-horizon limit of \eqref{eq: extremal black brane} and \eqref{eq: black brane field strength} with $n=d+1$) as background solution to the action \eqref{eq: action} and consider their perturbations.

The equations governing the dynamics of such perturbations $h \equiv \delta g$ and $\delta F$ are derived by perturbing \eqref{eq: background eqs}, which gives
\begin{subequations}
	\label{eq: perturbation equations}
	\begin{align}
		&\begin{aligned}
			0 = \delta M_{A_{2} \dots A_{d+1}}& \equiv \nabla^{A_{1}}\delta F_{A_{1}\dots A_{d+1}} - \frac{1}{2} F_{C A_{2} \dots A_{d+1}} \(2\nabla_{A_{1}} h^{A_{1} C} - \nabla^{C}h\) \\
			& - F_{A_{1} C A_{3} \dots A_{d+1}} \nabla^{[A_{1}} h\indices{^{C]}_{A_{2}}} - \dots - F_{A_{1} \dots A_{d} C} \nabla^{[A_{1}} h\indices{^{C]}_{A_{d+1}}}, \\
		\end{aligned} \\
		&\begin{aligned}
			0 = \delta E_{AB}&\equiv \delta G_{AB} - \frac{1}{2d!}\[2 \delta F_{A A_{2}\dots A_{d+1}} F\indices{_{B}^{A_{2}\dots A_{d+1}}} - d F_{A A_{2} A_{3} \dots A_{d+1}} F\indices{_{B B_{2}}^{A_{3}\dots A_{d+1}}} h^{A_{2}B_{2}} \] \\
			&+\frac{1}{4(d+1)!}\[h_{AB}F^{2} + g_{AB}\(2 \delta F_{C_{1}\dots C_{d+1}}F^{C_{1}\dots C_{d+1}} - (d+1)F_{C C_{2}\dots C_{d+1}} F\indices{_{D}^{C_{2} \dots C_{d+1}}} h^{CD}\)\], \\
		\end{aligned}
	\end{align}
\end{subequations}
with 
\begin{equation}
	\begin{aligned}
		\delta G_{AB} =& \nabla_{(A|}\nabla_{C}h\indices{^{C}_{|B)}} - \frac{1}{2}\nabla_{B}\nabla_{A}h - \frac{1}{2}g_{AB}\nabla_{C}\nabla_{D}h^{CD} - \frac{1}{2}\Box h_{AB} + \frac{1}{2} g_{AB}\Box h \\
		&+ \frac{1}{2} g_{AB} R_{CD}h^{CD} -\frac{1}{2}Rh_{AB} + h\indices{^{C}_{(A}}R_{B)C} - h\indices{_{C}_{D}}R\indices{_{A}^{C}_{B}^{D}} .
	\end{aligned}
\end{equation}
Our strategy is as follows. 
We will perform the KK reduction at the level of the equation of motion\footnote{Note that this is not the approach taken in \cite{Brown:2013mwa, Hinterbichler:2013kwa}, where the reduction is performed at the level of the action.}.
For the gravitational perturbations, we employ the decomposition introduced in \cite{Kodama:2000fa, Kodama:2003jz, Kodama:2003kk} to find decoupled scalar equations of motion to isolate the physical dynamical modes.
For the $d$-forms, it is useful to use the Hodge decomposition theorem to parameterise the gauge field/field strength in terms of co-exact forms, for which the eigenvalues of the Laplacian are known \cite{Camporesi:1994, Copeland:1984qk, Hinterbichler:2013kwa, Charalambous:2024tdj}.
In principle, there is a tower of these --- however, we are only interested in the ones which mix with the gravitational perturbations.
The other will decouple and act like scalars propagating on the fixed background geometry. 
Together, these will organise into tensor, vector, and scalar modes.
Their equations of motion will be of the form \eqref{eq: generic ads wave eq} \textit{i.e.} $\mathrm{AdS}_{d+1}$ wave equations, with an effective masses $\meff$ depending on the modes labelled by eigenvalues of the Laplacian on $S^{D-p-2}$ denoted by $\hat{\Delta}_{D-p-2}$. 
For the purposes of computing the KK spectrum of perturbations, we may choose to ignore all $x^{\alpha}$ dependence in our fields, effectively setting $\pp_{\alpha} =0$. 
We can do this without loss of generality at a technical level, as derivatives in the worldvolume directions $\pp_{\alpha}$ are necessarily packaged within $\Box_{\mathrm{AdS}}$ in order to be compatible with $\mathrm{ISO}(1,d-1)$ invariance.
By comparison with \eqref{eq: bg scalar eom}, we see that tracking the $\pp_{\rho}$ terms uniquely fixes the mass.

\subsubsection{Tensor Modes}

Let us start by considering tensor perturbations.
The tensor harmonics $\mathbb{T}_{ij}$ satisfy 
\begin{equation}
	\[\hat{\Delta}_{D-p-2}+\(\ell(\ell+\tilde{d})-2\)\]\mathbb{T}_{ij} = 0,\quad \mathbb{T}\indices{^{i}_{i}} = 0, \quad \hat{\DD}_{i}\mathbb{T}\indices{^{i}_{j}} = 0,
\end{equation}
where $k_{T}^{2} = \ell(\ell+\tilde{d}) - 2$, and the tensor perturbations are parameterised as
\begin{subequations}
	\begin{align}
		& \delta F =0 \\
		& \delta g = h_{T}(\rho) \mathbb{T}_{ij} \dd x^{i} \dd x^{j}.
	\end{align}
\end{subequations}
Maxwell's equation is trivially satisfied, and the remaining non-trivial equation of motion is
\begin{equation}
	-\frac{1}{2}\Box h_{ij} + \[\(\frac{1}{\tilde{d}+1} + \frac{1}{\tilde{d}(\tilde{d}+1)}-\frac{1}{2}\){}^{(\tilde{d}+1)}R - \frac{1}{2}{}^{(d+1)}R - \frac{1}{4}f^{2}\]h_{ij}=0, 
\end{equation}
which takes the form of \eqref{eq: bg scalar eom} with
\begin{equation}
	\meff^{2}L^{2} = \left(\frac{d}{\tilde{d}}\right)^{2} \ell(\ell+\tilde{d}).
\end{equation}
Note that this is precisely the effective mass of a probe scalar on the background geometry.

\subsubsection{Vector Modes}

Next, let us consider vector perturbations.
The vector harmonics $\mathbb{V}_{i}$ satisfy
\begin{equation}
	\[\hat{\Delta}_{D-p-2}+\(\ell(\ell+\tilde{d}) - 1\)\]\mathbb{V}_{i} = 0,\quad \hat{\DD}_{i}\mathbb{V}^{i} = 0,
\end{equation}
and we define from it the derived tensor harmonic
\begin{equation}
	\mathbb{V}_{ij} = -\frac{1}{k_{V}}\hat{\DD}_{(i}\mathbb{V}_{j)}.
\end{equation}
The vector modes are then parameterised as
\begin{subequations}
	\begin{align}
		&\delta F = \dd a \wedge \mathbb{V} + (-)^{d+1}a \wedge \dd \mathbb{V} \\
		& \delta g = 2f_{\mu}^{(V)}\mathbb{V}_{i} \dd x^{\mu} \dd x^{i}+ h_{V} \mathbb{V}_{ij} \dd x^{i} \dd x^{j},
	\end{align}
\end{subequations}
where we have written $\mathbb{V} = \mathbb{V}_{i} \dd x^{i}$ and $a \in \Omega^{d-1}(M)$, and $a$, $f_{\mu}^{(V)}$, and $h_{V}$ functions of $\rho$ only. 
With this parameterisation,
\begin{subequations}
	\begin{align}
		&\delta F_{\rho \alpha_{2}\dots \alpha_{d} i } = \pp_{\rho}a_{\alpha_{2}\dots \alpha_{d}} \mathbb{V}_{i} \\
		&\delta F_{\alpha_{2}\dots \alpha_{d} i j} = (-)^{d+1} a_{\alpha_{2} \dots \alpha_{d}} (\hat{\DD}_{i}\mathbb{V}_{j} - \hat{\DD}_{j} \mathbb{V}_{i}).
	\end{align}
\end{subequations}
We are ultimately interested in writing everything in terms of gauge-invariant variables, which are given by 
\begin{equation}
	f_{\rho} = -\frac{1}{2k_{V}} \pp_{\rho}h_{V}, \quad f_{\alpha} = \rho^{d+1}\pp_{\rho} \Omega_{\alpha}.
\end{equation}
It will also be convenient to define the Hodge dual of $a$ given by
\begin{equation}
	\tilde{a}_{\mu_{1} \mu_{2}} = \frac{1}{(d-1)!} \epsilon_{\mu_{1} \dots \mu_{d+1}}  a^{\mu_{3} \dots \mu_{d+1}}.
\end{equation}

Let us first consider Maxwell's equation. 
The first non-trivial component is 
\begin{equation}
	\delta M_{\rho \alpha_{3}\dots \alpha_{d} i} = \frac{(-)^{d-1}}{r_{H}^{2}} \gamma^{mn} \hat{\DD}_{m}\(\delta F_{\rho \alpha_{3} \dots \alpha_{d} n i}\) \propto a_{\rho \alpha_{3} \dots \alpha_{d}}
\end{equation}
for a generic (non-exceptional) $\ell$-mode, setting $a_{\rho \alpha_{3} \dots \alpha_{d}} =0$. 
This means that the only non-trivial component of $\tilde{a}$ is $\tilde{a}_{\rho\alpha}$.
The only other non-trivial components left is
\begin{equation}
	\begin{aligned}
		\delta M_{\alpha_{2} \dots \alpha_{d+1}} = \bigg[&- \frac{k_{V}^{2} + \tilde{d}}{r_{0}^{2}} a_{\alpha_{2} \dots \alpha_{d}} + \(\frac{\rho}{L}\)^{2}\(\pp_{\rho}^{2}a_{\alpha_{2} \dots \alpha_{d}} + \frac{-d+3}{\rho}\pp_{\rho}a_{\alpha_{2} \dots \alpha_{d}}\) \\
		&- F_{\rho \alpha_{2} \dots \alpha_{d+1}} \(\frac{\rho}{L}\)^{2} \(\pp_{\rho}f^{\alpha_{d+1}}+ \frac{2}{\rho}f^{\alpha_{d+1}}\)\bigg]\mathbb{V}_{i}.
	\end{aligned}
\end{equation}

Next, let us consider Einstein's equation.
Switching to the gauge-invariant variables automatically trivialises $\delta E_{\rho i}$ and $\delta E_{ij}$, and the only non-trivial equation left is
\begin{equation}
	\begin{aligned}
		\delta E_{\alpha i} &= \bigg\{-\frac{1}{2}\[\(\frac{\rho}{L}\)^{2} \pp_{\rho}^{2}f_{\alpha} + (d-1)\frac{\rho}{L^{2}} \pp_{\rho}f_{\alpha} + \(-\frac{k_{V}^{2}}{r_{0}^{2}} - \frac{d}{L^{2}}\)f_{\alpha}\] - \frac{1}{4}f^{2} f_{\alpha} \\
		&+ \frac{(-)^{d}}{2(d-1)!}f \epsilon\indices{^{\rho}_{\alpha}^{\alpha_{3}\dots \alpha_{d+1}}} \pp_{\rho}a_{\alpha_{3} \dots \alpha_{d+1}} \bigg\} \mathbb{V}_{i}.
	\end{aligned}	
\end{equation}

In the end, we are only left with solving 
\begin{subequations}
	\begin{align}
		0=&\(\frac{\rho}{L}\)^{2} \pp_{\rho}^{2}f_{\alpha} + (d-1)\frac{\rho}{L^{2}} \pp_{\rho}f_{\alpha} - \(\frac{d}{\tilde{d}}\)^{2} \frac{k_{V}^{2}-\tilde{d}}{L^{2}}f_{\alpha} - (-)^{d} f \(\pp_{\rho}\tilde{a}\indices{^{\rho}_{\alpha}}  + \frac{d-3}{\rho} \tilde{a}\indices{^{\rho}_{\alpha}}\) \\
		0=& \(\frac{\rho}{L}\)^{2} \pp_{\rho}^{2}\tilde{a}\indices{^{\rho}_{\alpha}} + (d-3) \frac{\rho}{L^{2}} \pp_{\rho}\tilde{a}\indices{^{\rho}_{\alpha}} - \frac{-3\tilde{d}^{2}+d \tilde{d}^{2}+ d^{2}\(k_{V}^{2}+\tilde{d}\)}{\tilde{d}L^{2}} \tilde{a}\indices{^{\rho}_{\alpha}} - (-)^{d} f \(\frac{\rho}{L}\)^{2} \pp_{\rho}f_{\alpha}
	\end{align}
\end{subequations}
which we can solve to give the effective mass
\begin{equation}
	\meff^{2}L^{2} = \frac{-d^{3} + d^{2} \(k_{V}^{2}+D-2\) +\tilde{d}^{2} \pm \sqrt{\tilde{d}} \sqrt{2d(D-2)k_{V}^{2}+\[(D-2)^{2}+d^{2}\]\tilde{d}}}{\tilde{d}^{2}}.
\end{equation}

\subsubsection{Scalar Modes}

Finally, let us consider the scalar perturbations.
The scalar harmonic $\mathbb{S}$ satisfies
\begin{equation}
	\(\hat{\Delta}_{D-p-2}+\ell(\ell+\tilde{d})\)\mathbb{S} = 0,
\end{equation}
and we define from it
\begin{equation}
	\mathbb{S}_{i}=-\frac{1}{k_{S}}\hat{\DD}_{i}\mathbb{S},\quad	\mathbb{S}_{ij} = \frac{1}{k_{S}^{2}}\hat{\DD}_{i}\hat{\DD}_{j}\mathbb{S}+\frac{1}{n}\gamma_{ij}\mathbb{S}.
\end{equation}
Then we parameterise the scalar modes as
\begin{subequations}
	\begin{align}
		&\delta F = \dd b\,  \mathbb{S} + (-)^{d}b \wedge \dd \mathbb{S} \\
		& \delta g = f^{(S)}_{\mu\nu} \mathbb{S}\, \dd x^{\mu} \dd x^{\nu} +  2 f_{\mu}^{(S)}\mathbb{S}_{i}\, \dd x^{\mu} \dd x^{i}+ \(h_{L}\mathbb{S} g_{ij} + h_{T}\mathbb{S}_{ij}\) \dd x^{i} \dd x^{j}
	\end{align}
\end{subequations}
for $b \in \Omega^{d}(M)$, and $b$, $f_{\mu\nu}^{(S)}$, $f_{\mu}^{(S)}$, $h_{L}$, and $h_{T}$ functions of $\rho$ only. 
With this parameterisation,
\begin{subequations}
	\begin{align}
		&\delta F_{\rho \alpha_{2}\dots \alpha_{d+1}} = \pp_{\rho}b_{\alpha_{2}\dots \alpha_{d+1}} \mathbb{S} \\
		&\delta F_{\alpha_{1}\dots \alpha_{d} i} = (-)^{d} b_{\alpha_{1} \dots \alpha_{d}} \hat{\DD}_{i}\mathbb{S}.
	\end{align}
\end{subequations}
In the end, we wish to write everything in terms of gauge-invariant variables, as in \cite{Kodama:2000fa}.
For the scalar perturbations, these are
\begin{subequations}
	\begin{align}
		&F = h_{L} + \frac{1}{\tilde{d}+1} \frac{1}{r_{0}^{2}} h_{T} \\
		&F_{\mu\nu} = f_{\mu\nu} + 2\DD_{(\mu}X_{\nu)},
	\end{align}
\end{subequations}
with
\begin{equation}
	X_{\mu} = \frac{1}{k_{S}}\(f_{\mu} + \frac{1}{k_{S}} \DD_{\mu}h_{T}\).
\end{equation}

Let us start by considering Maxwell's equations. 
The only non-trivial components of Maxwell's equations are 
\begin{subequations}
	\begin{align}
		\epsilon^{\rho \alpha_{2} \dots \alpha_{d+1}} \delta M_{\alpha_{2} \dots \alpha_{d+1}} \propto & \(\frac{\rho}{L}\)^{2}\(\pp_{\rho}^{2}\tilde{b}^{\rho} + \frac{d-1}{\rho}\pp_{\rho}\tilde{b}^{\rho} - \frac{d-1}{\rho^{2}} \tilde{b}^{\rho}\) - \frac{k_{S}^{2}}{r_{0}^{2}} \tilde{b}^{\rho} \\
		&\quad \quad - f \(\frac{\rho}{L}\)^{2}\( - \frac{1}{2} \pp_{\rho}f\indices{^{\lambda}_{\lambda}} + \frac{\tilde{d}+1}{2}\pp_{\rho}h_{L} - \frac{k_{S}}{r_{0}^{2}} f_{\rho}\) \\
		\epsilon^{\rho \alpha_{2} \dots \alpha_{d+1}}\delta M_{\rho \alpha_{3} \dots \alpha_{d+1}} \propto & \tilde{b}^{\alpha_{2}} - f f^{\alpha_{2}}.
	\end{align}
\end{subequations}
The latter is a constraint and fixes $\tilde{b}^{\alpha}$ in terms of gravitational perturbations.
In the decoupling limit, the former is a dynamical equation for $\tilde{b}^{\rho}$ --- however, these will be mixed with gravitational perturbations when $\mpl$ is finite.

All of the components of Einstein's equations are non-zero --- let us go through them systematically.
First of all, we note that
\begin{equation}
	\delta E\indices{^{\alpha}_{\beta}} \propto \(\frac{\rho}{L}\)^{2}\(\pp_{\rho}^{2}F\indices{^{\alpha}_{\beta}} + \frac{d+1}{\rho}\pp_{\rho}F\indices{^{\alpha}_{\beta}}\) - \frac{k_{S}^{2}}{r_{0}^{2}}F\indices{^{\alpha}_{\beta}},\quad \alpha \neq \beta.
\end{equation}
If we identify $\phi = f\indices{^{\alpha}_{\beta}}$ with $\alpha \neq \beta$, this is precisely \eqref{eq: bg scalar eom} with the effective mass of a scalar propagating on the fixed background.
Next, note that 
\begin{equation}
	\delta E_{\rho \alpha} \propto F_{\rho\alpha},
\end{equation}
which is just a constraint.
Together, $\delta M_{\rho \alpha_{3} \dots \alpha_{d+1}} =0$ and $\delta E_{\rho \alpha} =0$ imply $\delta E_{\alpha i} =0$.
With the change of variable $(h_{L},f_{\mu\nu}) \mapsto (F,F_{\mu\nu})$, the other equations still contain terms involving $f_{\mu}$ and derivatives thereof, and derivatives of $h_{T}$. 
We may use 
\begin{equation}
	\delta E_{\rho i} \propto -\frac{d(2\tilde{d}+d)}{2k_{S}\tilde{d}} \frac{1}{L^{2}}\pp_{\rho}h_{T} + \frac{k_{S}}{2} \pp_{\rho}\(F\indices{^{\gamma}_{\gamma}} + \tilde{d} F\) + \frac{k_{S}}{2\rho}\(F\indices{^{\gamma}_{\gamma}} - d F\indices{^{\rho}_{\rho}}\) + \frac{f k_{S}}{2}\tilde{b}_{\rho} - \frac{d}{\tilde{d}}\frac{D-2}{L^{2}}f_{\rho}
\end{equation}
to eliminate $h_{T}$ fully in terms of the other variables.
We can similarly use a combination of $\delta E_{ij}$ and $\delta E_{\rho\rho}$ to eliminate $f_{\rho}$ in favour of the other variables from the remaining equations of motion --- this leaves the orthogonal combination of these equations, which we denote $\delta \tilde{E}_{\rho\rho}$. 
Now, $\delta \tilde{E}_{\rho\rho}$ and the trace of $\delta E_{\alpha\beta}$ can be used to eliminate $F_{\rho\rho}$ --- both equations are linear in $\pp_{\rho}^{2}F\indices{^{\gamma}_{\gamma}}$ and $\pp_{\rho}^{3}F\indices{^{\gamma}_{\gamma}}$, so we may use them together to find relations of the form
\begin{subequations}
	\begin{align}
		&\pp_{\rho}^{3}F\indices{^{\gamma}_{\gamma}} = L_{1}\[F\indices{^{\gamma}_{\gamma}}, \pp_{\rho} F\indices{^{\gamma}_{\gamma}} ; F, \pp_{\rho}F, \pp_{\rho}^{2}F, \pp_{\rho}^{3}F ; F_{\rho\rho}, \pp_{\rho} F_{\rho\rho}, \pp_{\rho}^{2}F_{\rho\rho}\], \\
		&\pp_{\rho}^{2}F\indices{^{\gamma}_{\gamma}} = L_{2}\[F\indices{^{\gamma}_{\gamma}}, \pp_{\rho} F\indices{^{\gamma}_{\gamma}}; F, \pp_{\rho}F, \pp_{\rho}^{2}F; F_{\rho\rho}, \pp_{\rho} F_{\rho\rho}\].
	\end{align}
\end{subequations}
The consistency condition $\pp_{\rho}L_{2} =L_{1}$ is then an equation for $F_{\rho\rho}$ in terms of the other fields.
Using this consistency condition, we also find that
\begin{equation}
	L_{2} = \pp_{\rho}^{2}F\indices{^{\gamma}_{\gamma}} + \mathcal{E}\(F, \pp_{\rho}F, \pp_{\rho}^{2}F,\pp_{\rho}^{3}F\pp_{\rho}^{4}F\),
\end{equation}
so the equation for $\pp_{\rho}^{2}F\indices{^{\gamma}_{\gamma}}$ is actually the condition
\begin{equation}
	\begin{aligned}
		0 = \mathcal{E} &\equiv \pp_{\rho}^{4} F + \frac{2(d+3)}{\rho}\pp_{\rho}^{3}F - \[(d+1)(d-7)+ 2\frac{d^{2}k_{S}^{2}}{\tilde{d}^{2}}\]\frac{1}{\rho^{2}}\pp_{\rho}^{2}F \\
		&\quad - (d+1)\[ (2d+1)(d-1) + \frac{2d^{2}k_{S}^{2}}{\tilde{d}^{2}}\]\frac{1}{\rho^{3}}\pp_{\rho}F + \frac{k_{S}^{2}d^{4}}{\tilde{d}^{2}}\(-2 + \frac{k_{S}^{2}}{\tilde{d}^{2}}\)\frac{1}{\rho^{4}}F.
	\end{aligned}
\end{equation}
Using this, 
\begin{equation}
	\delta E_{\alpha\alpha} \propto \(\frac{\rho}{L}\)^{2}\(\pp_{\rho}^{2}\psi + \frac{d+1}{\rho} \pp_{\rho}\psi \) - \frac{k_{S}^{2}}{r_{0}^{2}}\psi,\quad \psi = F\indices{^{\gamma}_{\gamma}} - d F\indices{^{\alpha}_{\alpha}}\text{ (no sum)}.
\end{equation}
Once again, we see that this is \eqref{eq: bg scalar eom} when we identify $\psi = \phi$, and the effective mass is the same as the scalar propagating on the fixed background.
Further, the vanishing of $\mathcal{E}$ and its derivatives actually ensure that $\delta E_{ij}$ and $\delta M_{\alpha_{2}}$ vanish identically, so this is the desired decoupled equation of motion describing gravito-electromagnetic perturbations. 
Using a power law Ansatz, we can solve this and read off the effective masses for these perturbations to be
\begin{equation}
	\meff^{2}L^{2} = \frac{d^{2}}{\tilde{d}^{2}}\(k_{S}^{2} + \tilde{d}^{2} \pm \tilde{d} \sqrt{\tilde{d}^{2} + 4k_{S}^{2}}\).
\end{equation}

\subsection{Kaluza--Klein Spectrum \label{subsec: aretakis from kk}}

\begin{figure}
    \centering
    \includegraphics[width=0.45\textwidth]{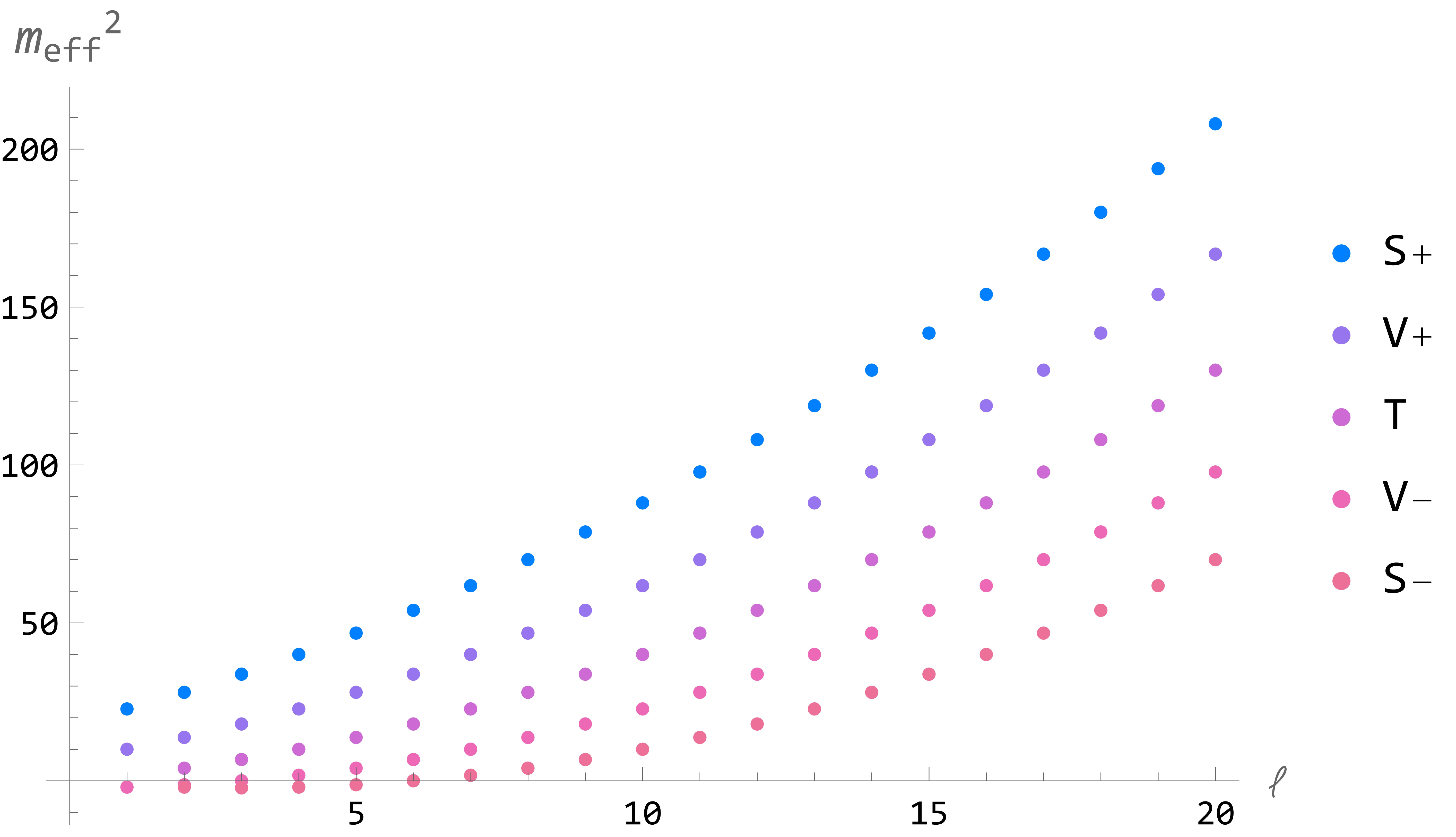}
    \hspace{0.05 \textwidth}
    \includegraphics[width=0.45\textwidth]{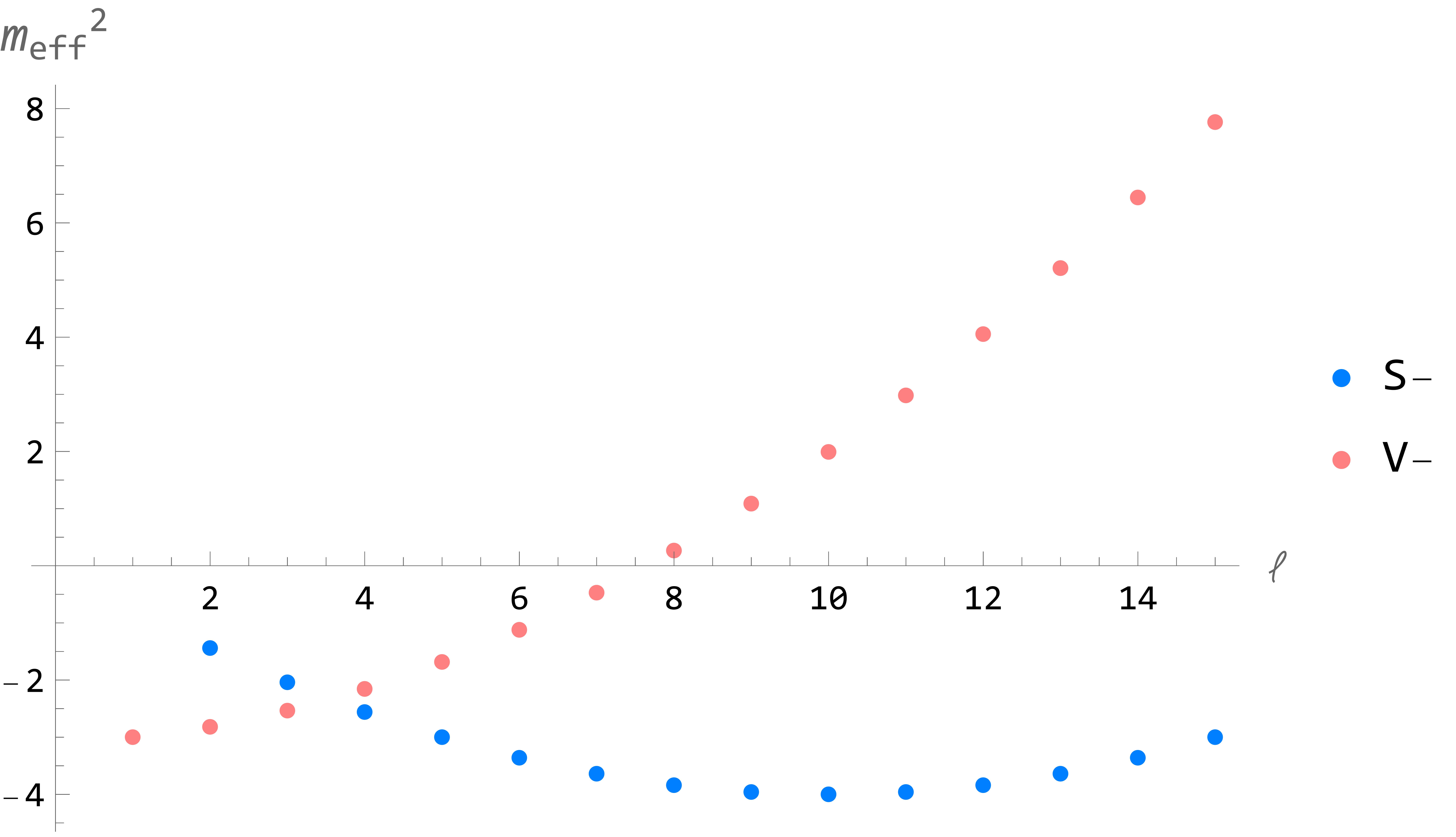}
    \caption{Plots for effective masses of different modes as a function of $\ell$ in units where $L=1$. 
    On the left, the spectrum of all (tensor, vector, and scalar) modes is shown for $(D,p) = (11,2)$. 
    It is clear that the inequalities \eqref{eq: mass hierarchy} hold.
    On the left, we have isolated gravitational scalar and vector modes for $(D,p) = (26,3)$ to illustrate that there is no hierarchy between gravitational scalar and vector modes which holds for all $\ell$.}
    \label{fig: spectroscopy}
\end{figure}

To summarise, this is the full spectrum of mixed gravitational and form field perturbations on the FR compactifications arising from electrically charged non-dilatonic extremal black branes:
\begin{subequations}
    \begin{flalign}
        \quad & \textbf{Rep.} && \textbf{Decoupled Mode} && \text{\textbf{Effective mass} $\meff^{2}L^{2}$} && \notag \\
        \hline \notag \\[-10pt]
        \quad & && h_{ij} && \notag \\
        \quad & \text{T} && F\indices{^{\alpha}_{\beta}},\, \alpha \neq \beta  && \left(\frac{d}{\tilde{d}}\right)^{2} \ell(\ell+\tilde{d}) \notag \\
        \quad & && F\indices{^{\gamma}_{\gamma}} - d F\indices{^{\alpha}_{\alpha}},\, \text{no sum} && \notag \\
        \quad & \text{V} && f_{\alpha} \quad \text{\&} \quad \tilde{a}\indices{^{\rho}_{\alpha}} && \frac{d^{2} \(\ell+1\)\(\ell+\tilde{d}-1\)+\tilde{d}^{2} \pm \sqrt{\tilde{d}} \sqrt{2d(d+\tilde{d})\[\ell(\ell+\tilde{d})-1\]+\[(d+\tilde{d})^{2}+d^{2}\]\tilde{d}}}{\tilde{d}^{2}} \notag \\
        \quad & \text{S} && F && \left(\frac{d}{\tilde{d}}\right)^{2}\[\ell^{2} + \tilde{d}\ell+\tilde{d}^{2} \pm \(2\ell+\tilde{d}\)\]. \notag
    \end{flalign}
\end{subequations}
There are two linearly independent vector and scalar perturbations, which reduce to purely gravitational or $d$-form perturbations in the decoupling limit respectively.
There were also perturbations written in terms of the scalar spherical harmonics $\mathbb{S}$, but behaved as \textit{scalar fields propagating on the fixed background} --- we have lumped these in with the tensor modes\footnote{These are not present when $p=0$ as the index $\alpha$ can only take $d$ distinct values.}.
The other effective masses agree with the near-horizon limits of the effective potentials given in \cite{Kodama:2003kk} for $p=0$ (and hence the scaling exponents computed in \cite{Chen:2024sgx}).
Some examples of KK spectra are shown in figure \ref{fig: spectroscopy}.

Note that for all physical values of $(d,\tilde{d},\ell)$, we have the following inequalities
\begin{equation}
    m^{2}_{S-/V-} < m^{2}_{T} < m^{2}_{S+/V+}, \quad m^{2}_{T/V+/S+} > 0.
    \label{eq: mass hierarchy}
\end{equation}
There is no relation between $m_{S-}^{2}$ and $m_{V-}^{2}$ which holds for all values of $(d,\tilde{d},\ell)$.
However, gravitational scalar and vector modes can give rise to massless modes for
\begin{equation}
    \ell = \tilde{d},\quad \ell= \frac{1}{2}\(-\tilde{d} + \sqrt{4 + \frac{\tilde{d}^{2}}{d^{2}}\[(d+4)d-4\]+ 4\frac{\tilde{d}}{d^{3/2}}\sqrt{d^{3}+2(d-1)d\tilde{d}+2(d-1)\tilde{d}^{2}}}\),
    \label{eq: massless deformations}
\end{equation}
respectively (when these are integers).
The examples in figure \ref{fig: spectroscopy} demonstrate these features.
For real scaling dimensions, fields on $\mathrm{AdS}_{d+1}$ should not lie below the BF bound $m_{\text{BF}}^{2}L^{2} = -d^{2}/4$ \cite{Breitenlohner:1982bm}.
It turns out that $m_{S/V-}^{2} \geq m_{\text{BF}}^{2}$, with the inequality saturated for
\begin{equation}
    \ell = \frac{\tilde{d}}{2}
    \label{eq: bf saturating modes}
\end{equation}
for gravitational scalar modes.
This means that all of the masses have masses lying above the BF bound.
This is perhaps not surprising --- some of these FR spaces arise as the near-horizon geometry of BPS branes, for which a positive-energy theorem exists.
While this alone is not sufficient to imply non-linear stability, it is certainly a good indicator.

Let us comment on electromagnetic duality.
As mentioned before, the action of free form fields is invariant under electromagnetic duality at the level of the background --- this maps $(F,d) \mapsto (\star F,\tilde{d})$.
\textit{E.g.} the spectrum for scalar modes on the electric background maps to the magnetic version under this map (\textit{cf.} the references above).
This is however more subtle for the other modes.
So far, by focusing on ones which mix with gravitational perturbations, we only considered $p$-form perturbations which fall into the SVT decomposition, but as mentioned previously this does not exhaust the full Hodge decomposition into co-exact forms \cite{Copeland:1984qk, Camporesi:1994, Yoshida:2019tvk, Charalambous:2024tdj}.
On a magnetically charged background, these will mix with the gravitational perturbations. 
By picking an electric/magnetic background, we have spontaneously broken the duality, and different perturbations will couple to gravity.
We therefore do not expect a duality mapping for the spectrum to hold with respect to Hodge duality on the full spacetime.
However, comparing to \textit{e.g.} \cite{Hinterbichler:2013kwa}, it is clear the KK spectrum for electrically charged extremal black branes is qualitatively and quantitatively similar to the ones for their magnetically charged analogues.

\subsubsection{Aretakis Instability}

What does this tell us about the Aretakis instability?

Generically, we have $\Delta \geq d/2$, with the bound being saturated by masses saturating the BF bound.
The coupled gravitational and $p$-form perturbations all satisfy this bound, with some gravitational scalar and vector modes saturating the bound. Therefore taking $\lceil d/2 \rceil$ (or $\lfloor d/2 + 1\rfloor$) transverse derivatives of the perturbations is sufficient to see the non-decay (or blow-up) of the field at the horizon.
This confirms that extremal black branes suffer from the Aretakis instability even in the absence of any additional fields, as perturbations of the fields which source the background geometry itself are already subject to it. 
In contrast to extremal black holes, extremal black branes seem to suffer less from the Aretakis instability.
Even when effective masses of the modes reduced onto $\mathrm{AdS}_{d+1}$ are comparable, more derivatives are needed to see non-decay of fields on extremal black branes. 

\section{Deformations of Extremal Black Branes \label{sec: deformations of branes}}

The scaling dimensions computed above also allow us to generalise the discussions in \cite{Horowitz:2022mly, Horowitz:2022leb, Horowitz:2024kcx} to extremal black branes. 
In section \ref{subsec: deformations branes subtleties} we will briefly comment on how some of the subtleties regarding deformations of black holes generalise to black branes.
We will then discuss the implications of the KK spectra for the gravitational/form field perturbations computed in section \ref{sec: perturbing freund-rubin} on these deformations in \ref{subsec: deformations results}.

As we saw, the near-horizon limit of the equation for perturbations of an extremal black brane geometry generically looks like an $\mathrm{AdS}_{d+1}$ wave equation \eqref{eq: bg scalar eom} with a mode-dependent effective mass.
Restricting ourselves to static and homogeneous deformations, \textit{i.e.} for which $\pp_{\alpha}\phi = 0$, the solutions to the equation of motion can be found to be
\begin{equation}
	\tilde{\phi} = c_{-} \rho^{\Delta^{(p+2)}-d} + c_{+} \rho^{-\Delta^{(p+2)}},
	\label{eq: scaling ansatz}
\end{equation}
where $\Delta^{(p+2)}$ are the $\mathrm{AdS}_{p+2}$ scaling dimensions in \eqref{eq: scaling dimension}.
Physically, we interpret these as the near-horizon limit of certain subclass of deformations of the full black brane geometry caused by some external sources.
The values of $\Delta^{(p+2)}$ then determine the behaviour of these deformations near the horizon.
Since $- \Delta^{(p+2)} < 0$ always, we ought to use boundary conditions at the horizon $\rho = 0$ to switch this solution off.
We are left with a solution scaling with a power $\gamma \equiv \Delta^{(p+2)} - d$, and it is not guaranteed that this is positive.
This is a direct generalisation of the discussions in \cite{Horowitz:2022mly, Horowitz:2022leb, Horowitz:2023xyl, Chen:2024sgx}.
This scaling exponent $\gamma$ then determines the regularity of the horizon for the deformed solutions.

\subsection{The Near-Horizon, Static, and Extremal Limits \label{subsec: deformations branes subtleties}}

There are several subtleties we should comment on.

One thing to note is that $\rho$ is not a good coordinate on the horizon --- what is to say that the scaling in $\rho$ is physical?
In the case of extremal black holes (with near-horizon geometry $\mathrm{AdS}_{2} \times S^{D-2}$), \cite{Horowitz:2022mly, Horowitz:2022leb} argued that $\rho$ was a good coordinate since it is generated by tangents to ingoing null geodesics.
In our case, this is suitably generalised by Gaussian null coordinates \eqref{eq: gaussian null}.
We see that, for a fixed point on the worldvolume, the scaling in the Gaussian null coordinate $\lambda$ in \eqref{eq: gaussian null} parameterising the distance away from the worldvolume is the same as in $\rho$.

It is also curious that we have restricted ourselves to solutions with worldvolume momentum $k_{\alpha} =0 $.
This is admittedly a fine-tuned case --- solutions with non-null $k_{\alpha}$ behave qualitatively different, and the solutions behave differently in the $k_{\alpha} \rightarrow 0$ limit.
From the out-set, we are therefore restricting ourselves to a very specific subset of perturbations --- this makes the set-up very finely tuned within the space of all possible perturbations to the original extremal black brane.
This is however well-motivated.
Even though we are just considering GR, we generically expect higher-derivative corrections to this --- in working with the two-derivative theory, we are therefore explicitly ignoring said EFT corrections.
In \cite{Chen:2024sgx} it was shown that the expansion in higher-derivative corrections breaks down when $k \gtrsim \Lambda^{2} \rho$.
Therefore, as we approach the horizon, we need $k^{2} \rightarrow 0$.
Poincar\'e invariance on the worldvolume means that $k_{\alpha}$ only shows up in $k^{2}$, so the condition $k_{\alpha} = 0$ is necessary and sufficient.
Therefore, even if we shall not consider these deformations as static limits of dynamical perturbations, restricting to this class of deformations is nonetheless a well-motivated condition.
For an explicit realisation of such a deformations, one may consider multi-centred black branes --- this is done in appendix \ref{app: multi-brane}.

Another subtlety is that we are allowing $\gamma<0$ while we used boundary conditions at the horizon to switch off the other branch of solutions.
To understand why we can accept the former in the extremal limit, we should appeal to the sub-extremal case.
Any pathology of the extremal solution that does not arise in the extremal limit of the corresponding sub-extremal solution should not be seen as physical \cite{Gubser:2000nd}.
In this case, the argument considering next-to-leading order terms in the extremal limit from \cite{Chen:2024sgx} follows through --- we show this explicitly in appendix \ref{app: extremal limit}.
Even solutions with $\gamma < 0$, which are singular on the horizon, arise from the extremal limit of sub-extremal solutions which are regular on the horizon.

\subsection{Results \label{subsec: deformations results}}

Generically, non-integer $\gamma$ indicate finite regularity of the metric at the horizon.
We find that generally $\gamma \geq -d/2$, which is saturated by masses saturating the BF bound.
Curvature invariants on the deformed geometry will scale as $\rho^{n\gamma}$ for $n\in \mathbb{N}_{>0}$, so any deformed solutions with $\gamma < 0$ will be so-called scalar polynomial (s.p.) singular. 
As discussed before, these arise as the extremal limit of non-extremal solutions which are regular at the horizon, and are therefore not to be dismissed from the outset.
There are also solutions which suffer from parallel-propagated (p.p.) singularities, \textit{i.e.} for which components of the Riemann tensor and hence tidal forces diverge in an orthonormal frame.
These are deformations with $0 < \gamma < 2$, which are not s.p. singular.
These are associated to a breakdown of wordline effective theory \cite{Horowitz:2024dch}. 

In terms of the effective masses, s.p. singularities correspond to effective masses which are negative (and above the BF bound).
We saw above that this is indeed possible, and even saturated, for the gravitational scalar and vector modes --- in particular, these are deformations with \eqref{eq: bf saturating modes}.
The deformations which correspond to massless modes correspond to marginal deformations --- these are ones with \eqref{eq: massless deformations}.
There is a now series of new examples of such deformations.
As discussed in \cite{Horowitz:2023xyl, Chen:2024sgx}, these are particularly interesting, as they are in principle dominated by higher-derivative corrections if one focuses on s.p. singularities.  
There are also modes which lead to p.p. singularities on the horizon.
In principle, this is possible for all modes, but the conditions are most cleanly stated for the scalar and tensor modes. 
\textit{E.g.}, tensor and electromagnetic scalar deformations lead to p.p. singularities when $\ell d < 2 \tilde{d}$ and $(\ell+\tilde{d})d < 2 \tilde{d}$ respectively.

The scaling dimension we derived for the multi-centred black branes in appendix \ref{app: multi-brane} behave like tensor or electromagnetic scalar modes. 
It is not possible for these to lead to (scalar polynomial) s.p. singularities, but (parallel-propagated) p.p. singularities are certainly possible. 
The fact that the horizons of multi-centred black holes and black branes are not smooth in this sense was previously noted in \cite{Welch:1995dh, Candlish:2007fh, Kimura:2014uaa} and \cite{Gowdigere:2012kq, Gowdigere:2014aca, Gowdigere:2014cqa} respectively.

\section{Concluding Remarks \label{sec: conclusion}}

To summarise, we used symmetry arguments to show that the scaling \eqref{eq: aretakis scaling} of transverse derivative of fields holds on general non-dilatonic extremal black $p$-brane backgrounds, extending the discussion in \cite{Cvetic:2020axz}.
Determining the severity of the Aretakis instability is reduced to computing the Kaluza--Klein spectrum of the fields present in the theory compactified on the $S^{D-d-2}$ within the near-horizon geometry. 
We use this to show that non-dilatonic extremal black branes suffer from the Aretakis instability even in the absence of additional fields, by computing the spectrum of perturbations to electrically charged backgrounds.
Finally, we relate the effective masses/scaling dimensions of perturbations to the scaling of deformations to the near-horizon geometry of non-dilatonic extremal black branes \eqref{eq: extremal black brane}.
This generalises the discussion in \cite{Horowitz:2022mly} to extremal black branes.

An obvious generalisation would be to compute the KK spectra of perturbations of backgrounds with dilatons and fermions, and specifically investigate what happens when the backgrounds preserve a certain amount of supersymmetry.
\textit{E.g.} for $D=11$ supergravity compactified on $S^{7}$ (which would arise as a near-horizon geometry of an extremal black 2-brane) and $\mathcal{N}=2$, $D=10$ supergravity compactified on $S^{5}$ (which would arise as a near-horizon geometry of an extremal black 3-brane), the KK spectrum was computed in \cite{Sezgin:1983ik, Biran:1983iy} and \cite{Gunaydin:1984fk, Kim:1985ez} respectively.
In these cases, it seems that the spectra show no qualitatively surprising features when it comes to the Aretakis instability --- the masses are parametrically $\meff^{2}L^{2} \sim \mathcal{O}(1)$ and satisfy the BF bound.

We leave for future work an investigation of the non-linear properties of these perturbations numerically.
Similar to \cite{Horowitz:2022leb}, it would be interesting to numerically construct the non-linearly deformed extremal black branes and find a description of the holographic dual. 
In the spirit of \cite{Lucietti:2012xr,Gelles:2025gxi}, it would also be useful to confirm if the near-horizon analysis of the Aretakis instability can be matched to an instability on the full spacetime.

Another interesting problem is to find the dynamical endpoint of the Aretakis instability. 
Generic perturbations of extremal branes will presumably lead to sub-extremal branes at intermediate times which are then subject to a GL-type instability, \textit{i.e.} the end-state will likely be an array of black holes.
However, one may be able to find a class of fine-tuned initial perturbations for which the end state will be a dynamical extremal black brane, the analogue of the solution found in \cite{Murata:2013daa,Angelopoulos:2024yev} for perturbations of extremal RN black holes.

\paragraph{UV Sensitivity ---} For both the Aretakis instability and the deformations to the near-horizon geometries, it was the integer part of scaling dimensions which determined how severe the blow-up at late times or strong the singularity on the horizon was.
The discontinuity of the floor/ceiling function here should be alarming --- for precisely integer scaling dimensions, this means that arbitrarily small corrections can have large effects.
For the sake of clarity: For integer $\Delta \geq 0$ and arbitrarily small $\epsilon >0$, 
\begin{equation}
	\lfloor \Delta - \epsilon \rfloor = \Delta - 1, \quad \lceil \Delta + \epsilon \rceil = \Delta + 1.
\end{equation}
We do expect such small corrections, when viewing our two-derivative theory \eqref{eq: action} as the leading-order terms in a low-energy effective field theory for an arbitrary UV completion. A breakdown in effective field theory therefore requires a scenario where the two-derivative theory is in some sense marginal, such that arbitrarily small corrections can lead to a \textit{qualitative} change. 
This is precisely the situation for $(D,p)=(5,0)$ --- here, the Aretakis instability is dominated by the gravitational scalar mode with $\ell=2$, for which $\Delta = 1$ exactly.
For this mode, the first derivative will be constant on the horizon, and transverse derivatives thereof blow up.
Higher-derivative corrections can push this mode above and below $\Delta = 1$, but this is a qualitative change --- even for an arbitrarily small correction for which $\Delta < 1$, only one derivative would be needed to see blow-up in the transverse derivatives.
This would mean that \textit{e.g.} the stress tensor diverges at late times.
For $D=4$, the leading-order contribution has $\Delta = 2$, but effective field theory corrections vanish at least up to four-derivatives.  
In higher dimensions, the leading-order contribution has non-integer $1 > \Delta > 0$, so that $\lfloor \Delta \rfloor = 0$ regardless of EFT corrections.
This is precisely the marginal deformation of the extremal RN black hole identified in \cite{Horowitz:2022mly}.
In the language of \cite{Hadar:2017ven}, this would constitute a breakdown of effective field theory at the horizon of an extremal black hole.
We expect that the resolution for the analogous problem for the deformations of extremal black holes in \cite{Chen:2024sgx} will apply directly to this set-up with the Aretakis instability --- the non-linearities of the perturbations will become important before the corrections from the higher-order EFT operators can become large, and in this can be even embedded into a partial UV completion.

If it were possible to excite individual modes, there are also several other modes which have integer scaling dimensions and would exhibit UV sensitivity in the sense described above.
Fields with integer scaling dimensions frequently appear in flux compactifications in type IIA/IIB string theory and M-theory exhibiting scale separation \cite{Aharony:2008wz, Conlon:2021cjk, Apers:2022zjx, Apers:2022tfm, Ning:2022zqx, Andriot:2023fss, VanHemelryck:2025qok, Farakos:2025bwf} and are associated to fields with enhanced (higher-order) shift symmetries (sitting in the discrete series representation of the (A)dS isometry group) \cite{Bonifacio:2018zex, Blauvelt:2022wwa}.
Scale separation is a crucial property for effective lower-dimensional descriptions of higher-dimensional physics --- the fact fact that we have found that modes with integer scaling dimensions appear to be sensitive to UV physics (again, in the sense that precisely when metric perturbation theory goes out of control is sensitive to UV corrections) might have implications for such scale-separated compactifications containing fields with integer scaling dimensions. 

\section*{Acknowledgements}

CYRC would like to thank Claudia de Rham and Andrew J. Tolley for useful conversations and collaboration on related topics.

The research of CYRC was supported by the Imperial College London President's PhD Scholarship and National Taiwan University 113L400-1NTU OED Grant. 
ADK is supported by the STFC Consolidated Grant ST/X000931/1.

CYRC would further like to thank Queen Mary University London for its hospitality while some of this work was completed.

\bibliographystyle{JHEP}

\bibliography{Bibliography.bib}

\appendix

\section{Alternative Representation of Solutions to the Near-Horizon Wave Equation \label{app: alternative deriv}}

In this section we discuss an alternative way to represent solutions to the near-horizon wave equations in extremal RN black holes and extremal black branes, working directly in Gaussian null coordinates \eqref{eq: warped product}.

\subsection{Extremal Reissner-Nordstr\"om}

The near-horizon wave equation in extremal RN, written in ingoing Eddington-Finkelstein coordinates, is given by
\begin{equation}
    \left[\Box_{\mathrm{AdS}_2}-\ell(\ell+1)\right]\phi_\ell=2\partial_v \partial_\rho \phi_\ell+\partial_\rho (\rho^2 \partial_\rho \phi_\ell)-\ell(\ell+1)\phi_\ell=0.
    \label{eq:AdS_2_wave}
\end{equation}
In these coordinates there is a canonical basis for the Killing vector fields of $\mathrm{AdS}_2$ which differs from \eqref{eq: killing vectors global basis}:
\begin{equation}
    \tilde{L}_0=v\partial_v-\rho\partial_\rho, \qquad \tilde{L}_+=\partial_v, \qquad \tilde{L}_-=v^2\partial_v-2(\rho v+1)\partial_\rho,
\end{equation}
These vectors also form a representation of the conformal algebra $\mathfrak{sl}(2,\mathbb{R})$
\begin{equation}
    [\tilde{L}_+,\tilde{L}_-]=-2L_0, \qquad [\tilde{L}_\pm,\tilde{L}_0]=\mp \tilde{L}_\pm.
\end{equation}
%We also note for later convenience that $\overline{L_\pm}=-L_{\mp}$ where the overbar denotes complex conjugation. 

Following a similar strategy as before, we can solve \eqref{eq:AdS_2_wave} by constructing simultaneous eigenfunctions of ${\cal C}$ and $\tilde{L}_0$
\begin{subequations}
    \begin{align}
        & {\cal C}\phi_{\ell,h} =\ell (\ell+1) \phi_{\ell,h}, \label{eq: casimir_ev_2} \\
        & \tilde{L}_0\phi_{\ell,h} =h \phi_{\ell,h}. \label{eq: boost_ev}
    \end{align}
\end{subequations}
Note that the general solution to \eqref{eq: boost_ev} is
\begin{equation}
    \psi_{\ell,h}=v^{-h} F_{\ell,h}(v\rho) 
    \label{eq: boost}
\end{equation}
for some function $F_{\ell,h}$. 
It then follows from \eqref{eq: casimir_ev_2} that $F_{\ell,h}$ as a function of $ z  = v \rho$ satisfies the hypergeometric differential equation
\begin{equation}
    (z^2+2z)\,F_{\ell,h}''(z)+2(z+1-h)\,F'_{\ell,h}(z)-\ell(\ell+1)\,F_{\ell,h}(z)=0.
\end{equation}
We impose boundary conditions at the horizon and at right timelike infinity: In particular, we require that $F_{\ell,h}$ is bounded at $z=0$ and that it decays at $z\to\infty$. 
Existence of a solution satisfying these boundary conditions enforces $h\geq \ell+1$ in which case the unique solution can be written as
\begin{equation}
    F_{\ell,h}(z)=c_{\ell,h}\,z^{-\ell-1}\,{}_2F_1\left(h+\ell+1,\ell+1;2\ell+2;-\frac2z\right).
    \label{eq: mode function sol}
\end{equation}
The solutions $\psi_{\ell,h}$ with $h=\ell+1+k$, $k\in\mathbb{Z}_{\geq 0}$ form the discrete series representation $\mathscr{D}^-_\ell$ of $SL(2,\mathbb{R})$. 
The lowest-weight element for each $\ell$ is given by
\begin{equation}
    \psi_{\ell,\ell+1}(v,\rho)=c_{\ell,\ell+1}v^{-\ell-1}\,(v\rho+2)^{-\ell-1},
\end{equation}
which satisfies
\begin{equation}
    \tilde{L}_- \psi_{\ell,\ell+1}=0.
\end{equation}
The solutions $\psi_{\ell,\ell+n+1}$ with $n\in\mathbb{Z}^+$ can then be generated by ``ascending" from $\psi_{\ell,\ell+1}$, \textit{i.e.} by repeated actions of $L_+$:
\begin{equation}
    \psi_{\ell,\ell+n+1}=(\pounds_{\tilde{L}_+})^n\psi_{\ell,\ell+1}, \quad n\in \mathbb{Z}^+.
\end{equation}
The lowest weight element exhibits the slowest decay ($v^{-\ell-1}$) on the horizon at late times, the higher weight elements $\psi_{\ell,\ell+n+1}$ decay as $v^{-\ell-n-1}$.
Moreover, we also see (by differentiating \textit{w.r.t.} $\rho$ and evaluating at $\rho=0$) that 
\begin{equation}
    \partial_\rho^k\phi_{\ell,\ell+n+1}\big|_{\rho=0} \sim v^{k-n-\ell-1} = v^{k-n-\Delta^{(2)}},
\end{equation}
as expected.

\subsection{Extremal black branes}

Consider first the case when $p\geq 3$ --- we will describe in the end of the section how the analysis changes in the cases $p=1$ and $p=2$. 
The wave equation in \eqref{eq: gaussian null} can be written as
\begin{equation}
	\begin{aligned}
		\Box_g\phi=&\,\Box_{\text{AdS}_{p+2}}\phi+r_0^{-2} \hat{\Delta}_{D-p-2}\phi \nonumber \\
		=&\, \cosh^{-2}\frac{\eta}{L}\left[2\partial_{\hat r}\partial_v\phi+\partial_{\hat r}\left(\frac{\hat r^2}{L^2}\partial_{\hat r}\phi\right)\right]+\frac{1}{a(\eta)}\partial_\eta\left(a(\eta)\partial_\eta \phi\right) \nonumber \\
		&\,+\frac{1}{L^{2}\sinh^{2} \frac{\eta}{L}}\, \hat{\Delta}_{p-1}\phi+\frac{1}{r_0^{2}}\hat{\Delta}_{D-p-2}\phi \\
		=&\,0,
	\end{aligned}
\end{equation}
where 
\begin{equation}
    a(\eta)=\cosh^2\frac{\eta}{L} \,\sinh^{p-1} \frac{\eta}{L}. \nonumber
\end{equation}
In Poincar\'e coordinates, the canonical form of the Killing vector fields of AdS$_{p+2}$ is
\begin{equation}
	\begin{aligned}
		&\tilde{L}_0=x^a\partial_a-\rho\partial_\rho, \qquad \qquad \tilde{L}_{ab}=x_a\partial_b -x_b \partial_a, \\
		&\tilde{L}_{-,a}=\left(\frac{L^4}{\rho^2}+\eta_{bc}x^bx^c\right)\partial_a-2x_a x^b\partial_b+2x_a \rho \partial_\rho, \qquad  \tilde{L}_{+,a}=\partial_a.
	\end{aligned}
\end{equation}
These generators satisfy the commutation relations associated with the conformal algebra
\begin{equation}
\begin{aligned}
    &[\tilde{L}_{\pm,a},\tilde{L}_0]=\pm \tilde{L}_{\pm,a} \\
    &[\tilde{L}_{+,a},\tilde{L}_{-,b}]=-2(\tilde{L}_0 \eta_{ab}-\tilde{L}_{ab}).
\end{aligned}
\end{equation}
Now let us transform these generators to GNCs. In particular, we have
\begin{equation}
    \tilde{L}_0=v\partial_v -\hat r\partial_{\hat r}, \qquad \tilde{L}_{+,0}=\partial_v, \qquad \tilde{L}_{-,0}=\frac{v^2}{L^2} \partial_v-2\left(\frac{\hat rv}{L^2}+1\right)\partial_{\hat r},
\end{equation}
\textit{i.e.} the generators $\tilde{L}_0$, $\tilde{L}_{\pm,0}$ form an $\mathfrak{sl}(2,\mathbb{R})$ subalgebra. 
Let us look for simultaneous eigenfunctions of $\tilde{L}_0$, $\hat{\Delta}_{p-1}\equiv \tilde{L}_{ab}\tilde{L}^{ab}$ and ${\cal C}$:
\begin{subequations}
	\begin{align}
		&{\cal C}\,\phi_{h,\ell,\ell'}=\nu_{\ell'}^2\,\phi_{h,\ell,\ell'} \\
		& \hat{\Delta}_{p-1}\,\phi_{h,\ell,\ell'}=-\ell'(\ell'+p-2) \,\phi_{h,\ell,\ell'}  \\
		& L_0 \phi_{h,\ell,\ell'}=-h\, \phi_{h,\ell,\ell'}.
	\end{align}
\end{subequations}
It follows that 
\begin{equation}
    \phi=\sum\limits_{h,\Delta ,\ell, \ell'} v^{-h}\phi_{h,\Delta ,\ell, \ell'}(v\hat r)\chi_{h,\Delta ,\ell, \ell'}(\eta) Y_{\ell}(\theta)Y_{\ell'}(\theta'),
\end{equation}
with $\psi_{h,\Delta ,\ell, \ell'}\equiv v^{-h}\phi_{h,\Delta ,\ell, \ell'}(vr)$ satisfying the AdS$_2$ wave equation
\begin{equation}
    2\partial_{\hat r}\partial_v\psi_{h,\Delta ,\ell, \ell'}+\partial_{\hat r}\left(\frac{\hat r^2}{L^2}\partial_{\hat r}\psi_{h,\Delta ,\ell, \ell'}\right)-\Delta(\Delta-1)L^{-2}\,\psi_{h,\Delta ,\ell, \ell'}=0,
\end{equation}
and $\chi_{h,\Delta ,\ell, \ell'}$ satisfying the ODE
\begin{equation}
    \frac{1}{a(\eta)}\frac{\rm d}{{\rm d}\eta}\left(a(\eta)\frac{\rm d}{{\rm d}\eta} \chi(\eta)\right)+L^{-2}\left[\frac{\Delta(\Delta-1)}{\cosh^{2}\frac{\eta}{L}}-\frac{\mu_{\ell}^2}{\sinh^{2}\frac{\eta}{L}}-\nu_{\ell'}^2\right]\chi(\eta)=0,
\end{equation}
with $\mu_{\ell}^2=\ell(\ell+p-2)$. 
An explicit solution of this ODE can be written in terms of hypergeometric functions. 
Requiring decay as $\eta\to\infty$ and boundedness at $\eta=0$ selects
\begin{equation}
    \chi(\eta)= \frac{\tanh^{\ell}\frac{\eta}{L}}{\cosh^{\Delta_+^{(p+2)}(\nu_{\ell'})}\frac{\eta}{L}} \,_{2}F_1\left[a_+,b_+,c_+,\cosh^{-2}\frac{\eta}{L}\right] ,
\end{equation}
with
\begin{subequations}
	\begin{align}
		a_+&= \frac12\left(\Delta^{(p+2)}(\nu_{\ell'} )-\Delta+\ell\right) \\
		b_+&= \frac12(\Delta^{(p+2)}(\nu_{\ell'})+\Delta+\ell-1) \\
		c_+&= \Delta^{(p+2)}(\nu_{\ell'} )-\frac{p-1}{2}.
	\end{align}
\end{subequations}
At $\eta\to\infty$ the solution behaves as
\begin{equation}
    \chi(\eta)= \cosh^{-\Delta_+^{(p+2)}(\nu_{\ell} )}\frac{\eta}{L} \left[1+\frac{a_+ b_+}{c_+}\cosh^{-1}\frac{\eta}{L}+\ldots\right].
\end{equation}
For generic choices of $\Delta$ the solution behaves as $\chi(\eta) \sim \tanh^{-p} (\eta/L)$ near the origin and thus it diverges as $\eta\to 0$ (since $p>0$). 
To prevent this, one needs to choose $\Delta$ such that the hypergeometric series stop at finite order. 
This requires one of $a_+$ or $b_+$ to be a negative integer. 
Since $\Delta\geq 0$, we must have
\begin{equation}
    \Delta=\Delta_+^{(p+2)}(\nu_{\ell'})+\ell+2n, \qquad n\in \mathbb{Z}^+,
    \label{eq:app_brane_delta_quant}
\end{equation}
in agreement with \eqref{eq:brane_quant_cond}. 
For fixed $\ell'$, $\ell$, we can obtain the solution corresponding to $n=0$ by imposing a lowest-weight condition. 
From our previous results, we know that the first condition fixes $h=\Delta$ which is the slowest possible decay rate compatible with regularity on the horizon and fall-off as $r\to\infty$. 
In particular, 
\begin{equation}
    L_{-1,0}\phi_{h, \Delta,\ell',\ell}=0, \qquad (L_{-1,1}-i L_{-1,2}) \phi_{h, \Delta,\ell',\ell}=0, \qquad (L_{1k}-i L_{2k}) \phi_{h, \Delta,\ell',\ell}=0. \label{eq:gnc_lowest_weight}
\end{equation}
The solution obeying the first of these conditions is
\begin{equation}
    \psi_{\Delta, \Delta,\ell,\ell'}(v,\hat r)=c_{\Delta, \Delta,\ell,\ell'}\, v^{-\Delta}\,\left(2+\frac{v\hat r}{L^2}\right)^{-\Delta}.
\end{equation}
The third condition fixes the spherical quantum numbers $-\ell_1=\ell_2=\ldots=\ell_{p-1}=\ell$ and the corresponding spherical harmonic
\begin{equation}
    Y_{-\ell,\ell, \ldots \ell}(\theta_1,\ldots \theta_{p-1})=c_{-,\ell}\, e^{-i\,\ell \theta_1} \prod\limits_{2\leq j\leq p-1} \sin^{\ell} \theta_{j}=c_{-,\ell}\,(\hat x_1-i\, \hat x_2)^{\ell}.
\end{equation}
Regarding the second condition, the generators $L_{-,a}$ in GNCs read as
\begin{equation}
	\begin{aligned}
		L_{-,a}&=\hat{x}_a \tanh \frac{\eta}{L} \left[v^2\partial_v+L^2\left(1+\left(1+\frac{v\hat r}{L^2}\right)^2\right)\partial_{\hat r}\right] \nonumber \\
		&-\hat{x}_{a} Lv\left(2+\frac{v\hat r}{L^2}\right)\partial_\eta+\frac{\left(2+\frac{v \hat r}{L^2}\right)v}{\tanh \frac{\eta}{L}}\left(\hat{x}_a\hat{x}_b-\delta_{ab}\right)\partial_b.
	\end{aligned}
\end{equation}
Imposing $(L_{-1,1}-i L_{-1,2}) \phi_{\Delta, \Delta,\ell',\ell}=0$ translates to a first order ODE for $\chi$
\begin{equation}
    \chi'(\eta)+\left(\Delta \tanh\frac{\eta}{L}-\frac{\ell}{\tanh\frac{\eta}{L}}\right)\chi(\eta)=0,
\end{equation}
whose solution is simply given by
\begin{equation}
    \chi(\eta)=\frac{\sinh^{\ell}\frac{\eta}{L}}{\cosh^\Delta \frac{\eta}{L}}.
	\label{eq:chi_sol}
\end{equation}
Plugging this into the radial ODE then gives
\begin{equation}
    \left[\Delta^2-\Delta(2\ell+p+1)\Delta+\ell(\ell+p+1)-\nu_{\ell'}^2\right]\left(\sinh\frac{\eta}{L}\right)^{\ell}\,\left(\cosh \frac{\eta}{L}\right)^{-\Delta}=0,
\end{equation}
which is satisfied if and only if
\begin{equation}
    \Delta=\ell+\frac{p+1}{2}\pm\sqrt{\left(\frac{p+1}{2}\right)^2+\nu_{\ell'}^2}=\ell+\Delta^{(p+2)}_{\pm}(\nu_{\ell'}).
\end{equation}
Imposing decay of the solution along the worldvolume directions as $\eta\to\infty$ requires $\Delta>\ell$ (\textit{c.f.} \eqref{eq:chi_sol}) which uniquely selects solutions with $\Delta=\ell+\Delta^{(p+2)}_{+}(\nu_{\ell'})$.

To summarise, for fixed $\ell$, $\ell'$ we constructed the solution with the lowest weight, given by
\begin{align}
    \phi_{0\ell\ell'}(v,\hat r,\eta,\theta,\theta')=C_{0,\ell,\ell'}\, v^{-\ell-\Delta^{(p+2)}_{+}(\nu_{l'} )}\frac{\sinh^{\ell}\frac{\eta}{L} \,Y_{-\ell,\ell, \ldots \ell}(\theta)\,Y_{\ell_1', \ldots \ell_{D-p-2}'}(\theta')}{\left[\cosh\frac{\eta}{L}\,\left(2+\frac{v\hat r}{L^2}\right)\right]^{\ell+\Delta^{(p+2)}_{+}(\nu_{\ell'} )}}. \label{eq:base_mode_p}
\end{align}
The same argument applies to the cases with $p=1,2$, the only difference is that the third condition in \eqref{eq:gnc_lowest_weight} is not needed. In the $p=1$ case there are no angular coordinates $\theta$. In the $p=2$ case there is a single angular coordinate $\theta$, and the $\theta$-dependence of the solutions is simply $Y_\ell(\theta)\propto e^{i\ell\theta}$.

The lowest weight solution \eqref{eq:base_mode_p} recovers the late-time, near-horizon behaviour (to leading order) found in section \ref{subsec: extremal black branes aretakis}. 
The solutions corresponding to $n>0$ have a faster fall-off on the horizon $\sim v^{\Delta}$ with $\Delta$ given by \eqref{eq:app_brane_delta_quant}.

\section{Multi-Centred Black Branes \label{app: multi-brane}}

To illustrate that these deformations to the near-horizon geometry of extremal black branes can be realised, it is useful to consider multi-centred black brane solutions.
Suppose this has $N+1$ centres located at $\mathbf{y}_{i}\in \mathbb{R}^{\tilde{d}+2}$ for $i \in \{0,\dots, N\}$ respectively.
The metric then takes the form of \eqref{eq: extremal black brane} with 
\begin{equation}
	H\(\mathbf{y}\) = 1 + \sum_{i=0}^{N} \frac{M_{i}}{|\mathbf{y}-\mathbf{y}_{i}|^{\tilde{d}}}.
\end{equation}
We are free to pick $\mathbf{y}_{0} = 0$ as a reference and identify $M = M_{1} $.
Then, let us define 
\begin{equation}
	\hat{\rho}_{i} = |\mathbf{y}_{i}|,\quad |\mathbf{y}| = \hat{\rho},\quad \mathbf{y} \cdot \mathbf{y}_{i} = \hat{\rho}\, \hat{\rho}_{i} \cos \theta_{i}.
\end{equation}
With this, we have
\begin{equation}
	H(\hat{\rho}) = H_{0}(\hat{\rho}) \[1 + h(\hat{\rho})\]
\end{equation}
where
\begin{equation}
	H_{0}(\hat{\rho}) = 1 + \frac{M}{\hat{\rho}^{\tilde{d}}},\quad h(\hat{\rho}) = \sum_{i=1}^{N} \frac{M_{i}}{M + \hat{\rho}^{\tilde{d}}} \sum_{j=0}^{\infty} C_{j}^{(\tilde{d}/2)}\(\cos \theta_{i}\) \(\frac{\hat{\rho}}{\hat{\rho}_{i}}\)^{j+\tilde{d}},
\end{equation}
where $C_{j}^{\alpha}$ are Gegenbauer polynomials.
In the limit $\hat{\rho}/\hat{\rho}_{i} \ll 1$, \textit{i.e.} when the reference brane is far away from the others, the metric \eqref{eq: extremal black brane} takes the form
\begin{equation}
	\dd s^{2} = \(\frac{\rho}{L}\)^{2} \eta_{\alpha\beta}\dd x^{\alpha} \dd x^{\beta} + \(\frac{L}{\rho}\)^{2} \[1 + \tilde{d} h(\rho) + \frac{\tilde{d}^{2}}{d} \frac{\dd h(\rho)}{\dd \log \rho}\]\dd \rho^{2} + r_{0}^{2} \[1 + h(\rho) \]^{2\tilde{d}} \dd \Omega_{\tilde{d}+1}^{2},
\end{equation}
with
\begin{equation}
	h(\rho) \sim \sum_{i=1}^{N} \frac{M_{i}}{M} \, \sum_{j=0}^{\infty} C_{j}^{(\tilde{d}/2)}\(\cos \theta_{i}\)  \(\frac{\rho}{\rho_{i}}\)^{j d/\tilde{d}+d}.
\end{equation}
From this, we can read off that the branes centred at $\mathbf{y}_{i}$ for $i \in \{1,\dots,N\}$ lead to a static deformation of the reference brane centred at $\mathbf{y}=\mathbf{0}$ which scales as 
\begin{equation}
	h \sim \rho^{j d/\tilde{d}+d}.
\end{equation}
near the horizon in an orthonormal frame.

\section{The Extremal Limit \label{app: extremal limit}}

In this appendix, we will briefly discuss an extension of the argument in \cite{Chen:2024sgx} to explain how deformations to extremal black branes even with singular scaling near the horizon ($\gamma<0$) are physical, as they arise as the extremal limit of sub-extremal solutions which are regular on the horizon.

From \cite{Gibbons:1994vm}, we see that the non-extremal version of the metric \eqref{eq: extremal black brane schwarzschild} is 
\begin{equation}
	\dd s^{2} = f_{+}f_{-}^{\frac{2-d}{d}} \dd t^{2} + f_{-}^{2/d} \delta_{ab} \dd x^{a} \dd x^{b} + \frac{1}{f_{+}f_{-}} \dd r^{2} + r^{2} \dd \Omega^{2},\quad f_{\pm} = 1- \(\frac{r_{\pm}}{r}\)^{\tilde{d}}.
\end{equation}
The Klein-Gordon equation for a massless scalar field propagating on this fixed background is
\begin{equation}
	0 = \Box \phi = \pp_{\alpha}\pp^{\alpha} \phi + \frac{1}{r^{\tilde{d}+1}}\pp_{r} \(r^{\tilde{d}+1}f_{+}f_{-} \pp_{r}\phi\) + \frac{1}{r^{2}}\hat{\Delta}_{S} \phi.
\end{equation}
Now, first consider defining $\epsilon = 2\frac{r_{+}-r_{-}}{r_{+}+r_{-}}$, which parameterises how far the background solution is from extremality.
Even though $\rho$ is the physical coordinate parameterising the distance from the horizon (as this is the one picked out by Gaussian null coordinates), it turns out to be more convenient to expand in $\rho_{*} = r-r_{+}$ instead.
Then, for stationary modes, at leading order in both the near-extremal and near-horizon expansions, the wave equation is
\begin{equation}
	\pp_{\rho_{*}} \[\ \(\frac{\rho_{*}}{r_{+}} + \epsilon \)\frac{\rho_{*}}{r_{+}}\tilde{d}^{2}\pp_{\rho_{*}} \phi\] - \meff^{2} \phi = 0,
\end{equation}
again with $\meff^{2} r_{+}^{2} = k_{S}^{2}$ determined by the eigenvalues of the operator $\hat{\Delta}_{S}$.
Then, to leading order in both the near-horizon and extremal limits, the solutions to this are
\begin{equation}
	\phi \sim A P_{\alpha}(z) + B Q_{\alpha}(z),
\end{equation}
where $P$ and $Q$ are the Legendre functions of the first and second kind respectively, and
\begin{equation}
	\alpha=\frac{1}{2}\(-1 + \sqrt{1 + \frac{4\meff^{2}r_{+}^{2}}{\tilde{d}^{2}}}\),\quad z = 1 + \frac{\rho_{*}}{r_{+}\epsilon}.
\end{equation}
In the sub-extremal case (for finite $\epsilon$) we ought to pick $B=0$ for finite boundary conditions at $\rho_{*}=0$. 
Then, the extremal limit $\epsilon \rightarrow 0$ picks out 
\begin{equation}
	\phi \sim \rho_{*}^{\alpha}.
\end{equation}
However, recall that the actual near-horizon coordinate is $\rho \sim L \left(\tilde{d}\hat{\rho}/r_{0}\right)^{1/d}$ in the near-horizon limit, so this reproduces the correct scaling exponent $\gamma = \alpha d$.
We see that the extremal limit does indeed reproduce the scaling solutions we found in the strictly extremal limit, always discarding the branch of solutions with scaling exponent $-\Delta^{(p+2)}$ and picking out the $\gamma$-branch of the solutions.
We should therefore view these as being physical even when $\gamma<0$.

\end{document}